\documentclass[journal,draftcls,onecolumn,12pt,twoside]{IEEEtran}

\IEEEoverridecommandlockouts
\usepackage{color}

\usepackage{amsthm,amsmath,amssymb}
\usepackage{mathtools}

\usepackage{verbatim}

\usepackage{cite}

\usepackage{url}
\usepackage{cases}
\usepackage{stmaryrd}

\usepackage[inline]{enumitem}

\usepackage{graphicx}

\ifCLASSOPTIONcompsoc
  \usepackage[caption=false,font=normalsize,labelfont=sf,textfont=sf]{subfig}
\else
  \usepackage[caption=false,font=footnotesize]{subfig}
\fi

\usepackage[switch]{lineno}
\usepackage[named]{algo}
\usepackage[noend]{algpseudocode}
\usepackage{algorithm}

\usepackage[flushleft]{threeparttable} 
\usepackage{bm}
\newtheorem{theorem}{Theorem}

\newtheorem{lemma}{Lemma}
\newtheorem{remark}{Remark}
\newtheorem{proposition}[theorem]{Proposition}

\begin{document}
\title{RIS-Aided Wideband Holographic DFRC}

\author{Tong Wei,~\IEEEmembership{Student Member,~IEEE}, Linlong Wu,~\IEEEmembership{Member,~IEEE}, Kumar Vijay Mishra,~\IEEEmembership{Senior Member,~IEEE}, M. R. Bhavani Shankar,~\IEEEmembership{Senior Member,~IEEE}


\thanks{The authors are with the Interdisciplinary Centre for Security, Reliability and Trust (SnT), University of Luxembourg, Luxembourg City L-1855, Luxembourg. E-mail: \{tong.wei@, linlong.wu@, kumar.mishra@ext., bhavani.shankar@\}uni.lu.}
\thanks{This work was supported by the Luxembourg National Research Fund (FNR) through the SPRINGER C18/IS/12734677.}
\thanks{The conference precursor to this work has been accepted for publication at the 2023 IEEE International Conference on Acoustics, Speech and Signal Processing (ICASSP).}
}

\maketitle

\begin{abstract}
 To enable non-line-of-sight (NLoS) sensing and communications, dual-function radar-communications (DFRC) systems have recently proposed employing reconfigurable intelligent surface (RIS) as a reflector in wireless media. 
 However, in the dense environment and higher frequencies, severe propagation and attenuation losses are a hindrance for RIS-aided DFRC systems to utilize wideband processing. To this end, we propose equipping the transceivers with the reconfigurable holographic surface (RHS) that, different from RIS, is a metasurface with an embedded connected feed deployed at the transceiver for greater control of the \textit{radiation amplitude}. This surface is crucial for designing compact low-cost wideband wireless systems, wherein ultra-massive antenna arrays are required to compensate for the losses incurred by severe attenuation and diffraction. We consider a novel  wideband DFRC system equipped with an RHS at the transceiver and a RIS reflector in the channel. We jointly design the digital, holographic, and passive beamformers to maximize the radar signal-to-interference-plus-noise ratio (SINR) while ensuring the communications SINR among all users. The resulting nonconvex optimization problem involves maximin objective, constant modulus, and difference of convex constraints. We develop an alternating maximization method to decouple and iteratively solve these subproblems. Numerical experiments demonstrate that the proposed method achieves better radar performance than non-RIS, random-RHS, and randomly configured RIS-aided DFRC systems.
\end{abstract}


\IEEEpeerreviewmaketitle

\section{Introduction}
    Reconfigurable intelligent surfaces (RISs) have recently emerged as an enabling technology for future wireless systems. A RIS consists of several passive or near-passive sub-wavelength metasurface elements \cite{hodge2020intelligent}. Conventionally, wireless systems assume that the fading channel is uncontrollable and is a significant factor that limits the performance because of random signal reflections, diffraction, and scattering in the wireless environment \cite{mishra2022machine}. RIS overcomes these fading channel limitations through the ability of metasurfaces to manipulate electromagnetic waves for applications such as arbitrary aperture beamforming \cite{wu2019towards}, frequency selective and high-impedance surfaces \cite{sievenpiper1999high}, polarization conversion \cite{zhu2013linear}, leaky-wave antenna \cite{minatti2015modulated}, beam focusing \cite{mishra2018reconfigurable}, and holographic imaging \cite{glybovski2016metasurfaces}. 

    In general, RIS is deployed as a reflector in wireless media. By exploiting the non-line-of-sight (NLoS) paths, the RIS-aided sensing \cite{esmaeilbeig2022irs} and communications \cite{renzo2020smart,wu2019intelligent} systems extend their coverage \cite{buzzi2022foundations,tang2021wireless}, suppress interference \cite{wu2019intelligent}, and secure the information transfer \cite{mishra2022optm3sec}. There is a rich heritage of research on non-RIS-based NLoS radars (see, e.g.,~\cite{watson2019non} and the references therein); but these techniques generally require prior and accurate knowledge of the geometry of propagation environment. In contrast, RIS-aided sensing exploits the NLoS echoes to compensate for the LoS path loss \cite{esmaeilbeig2022irs,aubry2021reconfigurable}. In wireless communications, RIS has been shown to enhance the coverage area by reflecting the impinging signals and hence overcome the severe line-of-sight (LoS) attenuation or blockage between the base station (BS) and multiple users (MU) \cite{tang2021wireless}. For example, in \cite{abeywickrama2020intelligent}, RIS was employed to minimize the total transmit power while ensuring the signal-to-interference-plus-noise ratio (SINR) among all users. Further, both the active and largely passive beamformers which employ at BS and RIS, respectively, are able to improve the overall quality-of-service (QoS) \cite{Ur2021joint}.

    Recently, the investigations of RIS focus on enhancing the performance of dual-function radar-communications (DFRC) \cite{mishra2019toward,wei2022quantized}, wherein sensing and communications jointly utilize the spectral and hardware resources \cite{elbir2022rise,wei2022multiple}. A single-RIS-aided DFRC was proposed in \cite{jiang2022intelligent} to maximize the radar SNR while utilizing the reflecting surface to simultaneously facilitate the target detection and single-user communications which is considered as the radar-centric design.  This set-up was extended to wideband DFRC with multiple RISs in \cite{wei2022multiple}. In practice, the RIS phase shifts are not continuous but quantized. This issue has been analyzed for DFRC in \cite{wang2022jointwaveform,wei2022simultaneous}. A few other recent studies on communications-centric DFRC design, where the RIS facilitates in maximizing secrecy rates \cite{mishra2022optm3sec}. 

    Early investigations on RIS-aided DFRC focused on narrowband sub-6 GHz 
    frequencies. Lately, rapid developments have taken place at millimeter-wave (mmWave) 
    communications 
    and sensing \cite{mishra2019toward} 
    to develop short-range technologies that exploit the large operational bandwidth at mmWave. 
    This band is characterized by severe attenuation during signal propagation. To compensate for these losses, extremely dense antenna arrays comprising a massive number of antenna elements are employed. 
    To this end, as one of the representative metamaterial antennas, reconfigurable holographic surface (RHS) \cite{hwang2020binary} has been proposed to realize such large arrays.  
    
    Different from RIS, the RHS is embedded in a large number of metamaterial radiation elements which are connected with a radio-frequency (RF) chain and are generally integrated with the transceivers. The RHS radiation elements exploit the holographic interference principle \cite{sleasman2015waveguide} to control the \textit{radiation amplitude} of the incident electromagnetic waves while also leading to compact and lightweight transceiver hardware \cite{deng2021survey,deng2022reconfigurable}. 
    This low-cost amplitude-control beamforming design was first proposed for the conventional antenna under orthogonal frequency-division-multiplexing (OFDM) transmissions \cite{gholam2011beamforming}. Then, it was extended to the holographic scenario for swift radiation beam control \cite{johnson2015sidelobe}.

    Initial RHS investigations were limited to wireless communications applications for flexible beam steering \cite{di2021reconfigurable,zeng2022reconfigurable}. Recently, it has been demonstrated for DFRC, wherein the holographic beam is aligned toward the target to ensure the communications signal-to-interference-plus-noise ratio (SINR) meets the requirements \cite{zhang2022holographic}. However, even with improved beam control, the RHS-aided systems yield poor performance in the absence of a stable line-of-sight (LoS) link thereby precipitating the need to also employ an RIS \cite{dardari2021holographic,wan2021terahertz}. Previous research largely focused on RIS-assisted wireless solutions for narrowband signaling thereby leading to frequency-independent passive beamformers. However, future wireless systems are expected to scale up in the spectrum and, therefore, exploit wide bandwidths available at the higher frequencies \cite{mishra2019toward}. Narrowband beamforming techniques are not usable for such wideband systems, where the resulting beam-squint effect \cite{elbir2021terahertz} could no longer be ignored.

    To overcome the above-mentioned limitations, in this paper, we jointly exploit the advantages of both RIS and RHS in a wideband DFRC system. We deploy the passive RIS \cite{li2021lntelligent} in the channel as a reflector while equipping the RHS at BS as the transceiver. Then, we jointly design the digital, holographic, and passive beamformers for the digital DFRC, RHS transceiver, and RIS, respectively. Our objective is to maximize the worst-case radar SINR over all the target, while also ensuring a certain minimum SINR for the different communication users. The resulting optimization problem involves nonconvex quadratic constraint quartic programming (QCQP) with coupled variables. We solve this challenging problem by first decoupling it into several subproblems which are solved via an alternating optimization (AO) algorithm.

We summarize our main contributions in this paper as follows.\\
\textbf{1) Wideband RHS model with beam-squint:}  Different from previous works focused on narrowband RHS-assisted DFRC systems \cite{deng2022reconfigurable,zhang2022holographic}, in this paper, we propose a more comprehensive wideband RHS model with OFDM signaling. Our proposed model allows for varying  the digital beamforming on different subcarriers, thereby offsetting the beam squint effect. Meanwhile, RHS can adjust the beam by controlling the radiation amplitude of the input signal.   \\  
\textbf{2) RIS-aided DFRC with RHS:} Contrary to prior works, we consider simultaneously harnessing the benefits of both RIS and RHS for DFRC applications. This joint deployment is especially helpful at higher frequencies, where  
the channel is LoS-dominant and NLoS-assisted \cite{mishra2019toward,elbir2021terahertz,wei2022multiple}. Here, the RIS beamforming in the NLoS paths boosts the indirect echoes. Further, RHS transceivers have a small form-factor and are able to quickly shape the radiation beampattern to overcome the fast fading channel. However, this deployment scheme imposes a new challenge for jointly designing the passive and holographic beamformers.
  \\
\textbf{3) Joint digital, holographic, and passive beamformer design:} We design the digital, holographic, and passive beamforming, and the receive filter, simultaneously, to maximize the worst-case radar SINR accounting for all the targets while guaranteeing the communications SINR.  To this end, we develop an alternating optimization (AO) framework to tackle the resulting nonconvex maximin problem. We first utilize the generalized Rayleigh quotient (GRQ) method to obtain the closed-form solution for the receive filter design. Then, we combine the Dinkelbach and majorization-maximization (MM) algorithm to tackle
the digital and holographic beamforming design. Finally, the consensus
alternating direction of multipliers (C-ADMM) \cite{yang2020dual}  and Riemannian steepest decent (RSD) \cite{alhujaili2019transmit}  approaches  are jointly utilized to solve the phase-shift design problem approximately. \\
\textbf{4) Extensive performance evaluation:} Our theoretical analyses and experimental investigation of the proposed RIS-aided holographic DFRC reveal a trade-off between communications and radar performance. Numerical results are also provided to illustrate the superior performance of the proposed algorithm in terms of the
minimum radar SINR compared with the non-RIS, random-RIS, and random-RHS DFRC
systems. 

The remainder of this paper is organized as follows. In the next section, we introduce the signal model and problem formulation for a RIS-aided wideband DFRC system with RHS. 
In Section \uppercase\expandafter{\romannumeral3}, we develop our AO-based algorithm to tackle the formulated nonconvex maximin problem, in which the corresponding subproblems are solved iteratively. We evaluate our methods in Section \uppercase\expandafter{\romannumeral4} through extensive numerical examples. Finally, we conclude in  Section \uppercase\expandafter{\romannumeral5}.

\emph{Notations:} Throughout this paper, vectors and matrices are denoted by lower case boldface letter and upper case boldface letter, respectively. The notations $(\cdot)^T$, $(\cdot)^{\ast}$ and $(\cdot)^H$ denote the operations of transpose, conjugate, and Hermitian transpose, respectively; ${\bf I}_L$ and ${\bf 1}_L$ denote the $L\times L$ identity matrix and all-ones vector of length $L$, respectively; $\otimes$ is the Kronecker product; $\mathrm{vec}(\cdot)$ is the vectorization of its matrix argument; $\mathrm{diag}(\cdot)$ and $\mathrm{blkdiag}(\cdot)$ denote the diagonal and block diagonal matrix, respectively;  $\langle{\mathbf a}\rangle$ denotes the diagonalization of the vector ${\mathbf a}$; $\preceq$ denotes the componentwise inequality for vector comparison. $|\cdot|$, $\|\cdot\|_1$, $\|\cdot\|_2$ and $\|\cdot\|_F$ represent the magnitude, $\ell_1$-norm, $\ell_2$-norm, and Frobenius-norm, respectively; $(\cdot)^{(l)}$ denotes the value of the variable at the $l$-th outer iteration. 

\begin{figure*}[t]
\centering
\includegraphics[width=0.85\textwidth]{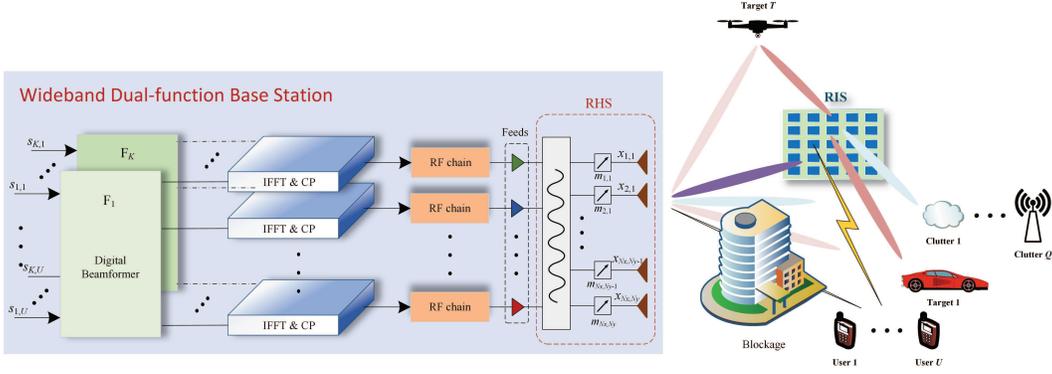}
\caption{Simplified illustration of RIS-aided wideband holographic DFRC system.}
\label{Fig1}
\end{figure*}

\section{System Model and Problem Formulation}
Consider a RIS-assisted wideband holographic DFRC system (Fig.~\ref{Fig1}). It employs orthogonal frequency-division multiplexing (OFDM) signaling with $K$ subcarriers. The far-field coverage area comprises $T$ radar targets, $Q$ clutter patches, and $U$ downlink users in a three-dimensional (3-D) Cartesian coordinate system. In particular, the dual-function base station (DFBS) is equipped with a RHS fed by $N_{RF}$ radio-frequency (RF) chains. The DFBS, (respectively RIS, user) is equipped with an uniform planar array (UPA) deploying $N_x^B$ ($N_x^R$, $N_x^U$) and $N_y^B$ ($N_y^R$, $N_y^U$) antenna elements with an inter-element spacing of $d_x^B$ ($d_x^R$, $d_x^U$) and $d_y^B$ ($d_y^R$, $d_y^U$) along the $x$- and $y$-axes, respectively. The resulting wideband space-frequency steering vectors of DFBS, RIS, and user are, respectively, given by
\begin{subequations} \label{ste_far}
 \begin{align}
 & {\bf a}_B(f_k,\theta,\psi) \!=\!\! 
       \left[{\bf a}^x_B(f_k,\theta,\psi)\!\otimes {\bf a}^y_B(f_k,\theta,\psi)
        \right],\\
 & {\bf a}_R(f_k,\theta,\psi) \!=\!\! 
       \left[{\bf a}^x_R(f_k,\theta,\psi)\!\otimes {\bf a}^y_R(f_k,\theta,\psi)
       \right], \\
 & {\bf a}_U(f_k,\theta,\psi) \!=\!\! 
       \left[
{\bf a}^x_R(f_k,\theta,\psi)\!\otimes {\bf a}^y_R(f_k,\theta,\psi)       \right],
 \end{align}
\end{subequations}
%
%
where $\theta\in[0,2\pi]$ ($\psi\in[0,\frac{\pi}{2}]$) are azimuth (elevation) angles; $\lambda_k=f_k/c$ and $f_k=f_c\!+\!\hat{f}_k$ are, respectively, wavelength and frequency of the $k$ subcarrier; $f_c$ and $\hat{f}_k =k \triangle f$ are the carrier frequency and $k$-th subcarrier offset frequency, respectively; $\triangle f$ is the subcarrier spacing and ${\bf a}^x_l(f_k,\theta,\psi)$ $\left({\bf a}^y_l(f_k,\theta,\psi)\right)$ denote the steering vectors along the $x$- ($y$-) axis as 
\begin{subequations} \label{ste_far2}
 \begin{align}
 & {\bf a}^x_l(f_k,\theta,\psi) = \left[ 1, e^{-j\frac{2\pi}{\lambda_k}2\mu_x^l},   \cdots,
e^{-j\frac{2\pi}{\lambda_k}(N_x^l\!-\!1)\mu_x^l} \right]^T, ~l\in\{B, R, U\},\\
 & {\bf a}^y_l(f_k,\theta,\psi) = \left[ 1, e^{-j\frac{2\pi}{\lambda_k}2\mu_y^l},   \cdots,
e^{-j\frac{2\pi}{\lambda_k}(N_y^l\!-\!1)\mu_x^l} \right]^T, ~l\in\{B, R, U\},
 \end{align}
\end{subequations}
where $\mu_x=d_x\cos{\theta}\cos{\psi}$, and $\mu_y=d_y\sin{\theta}\cos{\psi}$ are the direction cosines \cite{nai2010beampattern}.  To simply the notation, hereafter, we denote $N_B=N_x^B\!\times\!N_y^B$, $N_R=N_x^R\!\times\!N_y^R$, and $N_U=N_x^U\!\times\!N_y^U$ as the total number of array elements in the antennas of the DFBS, RIS and communication user.
Then, we denote the 3-D position vectors of RIS, $t$-th target, DFBS, $q$-th clutter, and $u$-th user, respectively, by
\begin{subequations} \label{position}
 \begin{align}
   &{\bf p}_R=[x_R,y_R,z_R],~{\bf p}_T(t) =[x_T(t),y_T(t),z_T(t)],  \\ 
   &{\bf p}_B=[x_B,y_B,z_B],~{\bf p}_C(q)=[x_C(q),y_C(q),z_C(q)],~{\bf p}_U(u)=[x_U(u),y_U(u),z_U(u)].
 \end{align}
\end{subequations}

Denote $\theta_{BR}$, $\theta_{Bu}$, $\theta_{Bt}$, and $\theta_{Bq}$  as the azimuth angles from DFBS to the RIS, $u$-th user, target, and $q$-th clutter, respectively ($\psi_{BR}$, $\psi_{Bu}$, $\psi_{Bt}$, $\psi_{Bq}$ for elevation). Further, designate the azimuth  angles from the RIS to DFBS, $u$-th user, target, and $q$-th clutter by $\theta_{RB}$, $\theta_{Ru}$, $\theta_{Rt}$, and $\theta_{Rq}$, respectively ($\psi_{RB}$, $\psi_{Ru}$, $\psi_{Rt}$, $\psi_{Rq}$ for elevation). Similarly,  $\theta_{BR}$, and $\theta_{Bq}$  are the azimuth  angles from $u$-th user to DFBS, and RIS, respectively ($\psi_{BR}$,  $\psi_{Bq}$ for elevation).
\subsection{RHS-Based Wideband Transmit Signal Model}
The RHS-based transceivers generate the emitted signal following the holographic interference principle. We refer the interested readers to \cite{di2021reconfigurable} (and references therein) for the details of the operational principle of holographic systems. Here, we follow the same model but adapt it for wideband DFRC.

\paragraph{OFDM Precoding} We denote the transmit symbol vector at the $k$-th subcarrier by 
${\bf s}_k=[s_{k,1},\cdots,s_{k,U}]^T\in\mathbb{C}^{U\times 1}$, with $\mathbb{E}\{{\bf s}_k{\bf s}_k^H\}={\bf I}_U$. Let ${\bf F}_k$ denote the frequency-dependent beamformer to enable multiuser (MU) communications and mitigate the beam-squint effect\cite{wang2019overview}. After the digital beamforming,  the frequency-domain signal at $k$-th subcarrier is
\begin{align}\label{sig_fre}
\widetilde{\bf x}_T[f_k]={\bf F}_k{\bf s}_k\in\mathbb{C}^{N_{RF}\times 1}, k=1,\cdots,K.
\end{align}
Further, $N_{RF}$ RF chains on the DFBS are connected to a RHS having $N_B$ discrete antenna elements\footnote{Note that the number of feeds should be greater or equal to the number of active data symbols in order to guarantee the decoder performance but less than the number of RHS element to reduce the hardware cost.}.
Applying $N_{RF}$ $K$-point inverse discrete Fourier transform (IDFT)  to \eqref{sig_fre} yields the baseband signal
\begin{align}\label{sig_time}
 {\bf x}_T(t)= {\sum}_{k=1}^{K}\widetilde{\bf x}_T[f_k]e^{\mathrm{j}2\pi{f_k}t} ={\sum}_{k=1}^{K}{\bf F}_k{\bf s}_k e^{\mathrm{j}2\pi{f_k}t},
\end{align}
where $t\in(0,T_s]$ and $T_s$ denotes the OFDM duration excluding the cyclic prefix (CP). 

\paragraph{RHS Beamforming} Following the model presented in \cite{di2021reconfigurable},  the electromagnetic response of the RHS at the $k$-th subcarrier takes the form
%
\begin{align}\label{RHS_resp}
 &{\bf V}[f_k]={\bf M}{\bf V}_k\in\mathbb{C}^{N_x^BN_y^B\times N_{RF}},
\end{align} 
where  the matrix ${\bf M}\!=\!\mathrm{diag}\left[m_{1,1},\cdots,\right.$ 
 $\left.m_{1, N_y^B},\cdots,m_{N_x^B,1},\cdots,m_{N_x^B,N_y^B}\right], 0\leq m_{x, y}\leq 1$ is the amplitude-control beamformer of $(x, y)$-th RHS element. Further, ${\bf V}_k(p,q)\!=\!e^{-2j\pi\gamma{D_{p,q}}/{\lambda_k}}$,  where $D_{p,q}$ denotes the distance between the $p$-th RHS element and $q$-th feed, $ p=1,\cdots,N_x^BN_y^B, q=1,\cdots,N_{RF}$, and $\gamma$ is the refractive index of the RHS material. Consequently,  the matrix ${\bf V}_k, k=1,\cdots,K$ is known if the structure of holographic surface is fixed.

\paragraph{DFBS transmit signal} After the holographic beamforming to \eqref{sig_time}, the transmitted signal in passband (excluding CP) takes the form
\begin{align}\label{trans}
\mathbf{x}(t) = {\bf M}{\sum}_{k=1}^{K}{\bf V}_k{\bf F}_k{\bf s}_k e^{\mathrm{j}2\pi {f_k}t}.
\end{align}
This signal is utilized to detect the targets and enable MU communications, simultaneously.  Meanwhile, for wideband
DFRC, the transmit power should meet the system requirement. In order to fully utilize the bandwidth, herein, we assume the transmit power satisfies
\begin{align}\label{power}
\|{\bf M}{\bf V}_k{\bf F}_k\|_F^2 \leq \mathcal{P}_k, \forall k
\end{align}
where $\mathcal{P}_k$ is the maximum  power assigned to the $k$-th subcarrier.
%
%
\subsection{Communications Receiver}
\label{ssec:Comm_Mod}
%

Denote the direct DFBS-user, RIS-user, and DFBS-RIS (in which only the LoS component is considered) channels at the $k$-th subcarrier frequency by ${\bf H}_{C_u,k}^{\mathrm{dir}}$, ${\bf H}_{C_u,k}^{\mathrm{RIS}}$, and ${\bf G}_k$, respectively. 
Following these prevalant channel models summarized in \cite{elbir2021terahertz} (and references therein), the aforementioned wideband channel components are given, respectively, by 
\begin{align}\label{comm_channel}
&{\bf H}_{C_u,k}^{\mathrm{dir}} \!=\! \sqrt{\frac{\Upsilon_{C_u}^{\mathrm{dir}}}{1\!+\!\Upsilon_{C_u}^{\mathrm{dir}}}}g_{Bu,k}{\bf a}_U(f_k,\theta_{uB},\psi_{uB}){\bf a}_B^T(f_k,\theta_{Bu},\psi_{Bu})\nonumber\\
&~\quad+\sqrt{\frac{1}{1+\Upsilon_{C_u,k}^{\mathrm{dir}}}}\sum_{l_d=1}^{L_d}g_{l_d,k}{\bf a}_U(f_k,\hat\theta_{l_d},\hat\psi_{l_d}){\bf a}_B^T(f_k,\theta_{l_d},\psi_{l_d}),   \\
&{\bf H}_{C_u,k}^{\mathrm{RIS}} \!=\! \sqrt{\frac{\Upsilon_{C_u}^{\mathrm{RIS}}}{1\!+\!\Upsilon_{C_u}^{\mathrm{RIS}}}}g_{Ru,k}{\bf a}_U(f_k,\theta_{uR},\psi_{uR}){\bf a}_R^T(f_k,\theta_{Ru},\psi_{Ru})\nonumber  \\
&~\quad+\sqrt{\frac{1}{1+\Upsilon_{C_u}^{\mathrm{RIS}}}}\sum_{l_r=1}^{L_r}g_{l_r,k}{\bf a}_U(f_k,\hat\theta_{l_r},\hat\psi_{l_r}){\bf a}_R^T(f_k,\theta_{l_r},\psi_{l_r}), \\
&\quad~{\bf G}_k = g_{BR,k}{\bf a}_R(f_k,\theta_{RB},\psi_{RB}){\bf a}_B^T(f_k,\theta_{BR},\psi_{BR}),
\end{align}
where $g_{\ast,k}=\sqrt{K_0(\frac{r_0}{r})^{\epsilon}}$ is the distance-dependent path loss, $K_0$ is the path loss at the $r_0$ reference distance, $r$ is the distance of the corresponding path, $L_d$ and $L_r$ denote the number of NLoS path for DFBS-user link and RIS-user link, ${\epsilon}$ is the path loss exponent (ranging from 2$-$4), and $\Upsilon^{\ast}_{C_u}$ denote the Rician factor for the corresponding path such that
\begin{equation}\label{block}
\left\{\begin{aligned}
     \Upsilon^{\ast}_{C_u}=0,  & \quad\mathrm{blockage},\\
     \Upsilon^{\ast}_{C_u}\neq0, & \quad\mathrm{LoS~path}.
     \end{aligned}
  \right.
\end{equation}
To simplify the design procedure, the aforementioned channels are estimated \textit{a priori} \cite{vlachos2019wideband,shtaiwi2021channel}.

%
Subsequent to baseband conversion, CP removal and $N_U$ $K$-point discrete Fourier transform (DFT), the received signal of $u$-th user on the $k$-th subcarrier is
\begin{align}\label{comm_rec}
{\bf y}_{C_u}[f_k]={\bf H}_{C_u,k}{\bf M}{\bf V}_k{\bf F}_k{\bf s}_k+{\bf n}_{C_u}[f_k] 
  = ({\bf H}_{C_u,k}^{\mathrm{dir}}\!+\!{\bf H}_{C_u,k}^{\mathrm{RIS}}{\bf\Phi}{\bf G}_k){\bf M}{\bf V}_k{\bf F}_k{\bf s}_k\!+\!{\bf n}_{C_u}[f_k],
\end{align}
where ${\bf H}_{C_u,k}$ denotes the composite channel between BS and $u$-th user, ${\bf\Phi}$ denotes the phase-shift matrix which is common across all subcarriers, 
${\bf n}_{C_u}[f_k]$ denotes the  zero mean white Gaussian noise  with  covariance $\sigma_r^2{\bf I}$ at the receiver front end for $k$-th subcarrier.

\begin{remark}
Due to the passive nature of RIS which precludes baseband signal processing unit, RIS can only work in the resonance frequency and thus the phase-shift is common for all subcarriers \cite{yang2020intelligent,zhang2021joint,zhang2020capacity}. Hence, the beam squint effect which degrades the performance of wideband system is also inevitable at RIS. While there exist some works aiming to design subcarrier specific  phase-shifts  to overcome the squint \cite{li2021lntelligent}, the complexity of their hardware implementation is beyond the envisaged DFRC system   \cite{wu2022wideband}. Hence, in this paper, we consider the fully passive RIS-assisted wideband DFRC with RHS. The active wideband RIS scenario will be explored in future work.    
\end{remark}
%
%
%
At the $k$-th subcarrier of $u$-th user, a digital combiner ${\bf w}_{k,u}$ is utilized to filter the received signal and estimate the transmitted symbol as 
\begin{align}\label{comm_filter}
\tilde{s}_{k,u} =& {\bf w}_{k,u}^H{\bf y}_{C_u}[f_k] \nonumber\\
                =&{\bf w}_{k,u}^H{\bf H}_{C_u,k}{\bf M}{\bf V}_k{\bf f}_{k,u}{ s}_{k,u} +{{\sum}_{i\neq u}^{U}{\bf w}_{k,u}^H{\bf H}_{C_u,k}{\bf M}{\bf V}_k{\bf f}_{k,i}{ s}_{k,i}}+{\bf w}_{k,u}^H{\bf n}_{C_u}[f_k].
\end{align} 
The quality of the estimate, ${s}_{k,u}$, is determined by the SINR, the typical metric for benchmarking the link performance of communication.
According to \eqref{comm_filter}, the SINR of $u$-th user on $k$-th subcarrier is 
\begin{align}\label{SINR_ck}
  \mathrm{SINR}_{C_u,k}  
    \!=\! \frac{\|{\bf w}_{k,u}^H{\bf H}_{C_u,k}{\bf M}{\bf V}_k{\bf F}_{k}{\bf\Lambda}_u\|^2}{\|{\bf w}_{k,u}^H{\bf H}_{C_u,k}{\bf M}{\bf V}_k{\bf F}_{k}\widetilde{\bf\Lambda}_u\|^2\!+\!\sigma_c^2{\bf w}_{k,u}^H{\bf w}_{k,u}}.    
\end{align}
Then, the average SINR over all subcarriers  for $u$-th user is 
\begin{align}\label{SINR_c_hat}
  \widetilde{\mathrm{SINR}}_{C_u}  
    \!=\!\frac{1}{K}{\sum_{k=1}^{K}}\frac{\|{\bf w}_{k,u}^H{\bf H}_{C_u,k}{\bf M}{\bf V}_k{\bf F}_{k}{\bf\Lambda}_u\|^2}{\|{\bf w}_{k,u}^H{\bf H}_{C_u,k}{\bf M}{\bf V}_k{\bf F}_{k}\widetilde{\bf\Lambda}_u\|^2\!+\!\sigma_c^2{\bf w}_{k,u}^H{\bf w}_{k,u}}
\end{align}
Note that \eqref{SINR_c_hat} is composed by the summation of a set of quartic fractional function in terms of receive filter and passive and holographic beamforming which is difficult to tackle. Hence, we reformulate \eqref{SINR_c_hat} as the sum-average SINR of $u$-th user \cite{xu2023bandwidth}:
\begin{align}\label{SINR_c}
  \mathrm{SINR}_{C_u}  
    \!=\! \frac{\frac{1}{K}{\sum_{k=1}^{K}}\|{\bf w}_{k,u}^H{\bf H}_{C_u,k}{\bf M}{\bf V}_k{\bf F}_{k}{\bf\Lambda}_u\|^2}{{\sum_{k=1}^{K}}(\|{\bf w}_{k,u}^H{\bf H}_{C_u,k}{\bf M}{\bf V}_k{\bf F}_{k}\widetilde{\bf\Lambda}_u\|^2\!+\!\sigma_c^2{\bf w}_{k,u}^H{\bf w}_{k,u})},    
\end{align}
where ${\bf\Lambda}_u$ denotes the selection matrix with $u$-th diagonal element
is one and the others are zero and $\widetilde{\bf\Lambda}_u\!=\!{\bf I}_U\!-\!{\bf\Lambda}_u$, the numerator and denominator of the right hand side of \eqref{SINR_c} denote the average desired signal power and total multi-user interference (MUI) plus noise power, respectively, for $u$-th user over all subcarriers. 

\begin{lemma} For any communication user, the sum-average SINR in \eqref{SINR_c} is a lower bound on the average SINR in \eqref{SINR_c_hat}, i.e. ${\mathrm{SINR}}_{C_u} \leq \widetilde{\mathrm{SINR}}_{C_u} $. 
\label{lem:SINR}
\end{lemma}
\begin{IEEEproof}
Let us first simplify the SINR in \eqref{SINR_c} and  \eqref{SINR_c_hat}  as 
\begin{subequations}\label{SINR_pro}
\begin{flalign}
    \mathrm{SINR}_{C}  & = \frac{1}{K}(\frac{a_1+a_2+\cdots+a_K}{b_1+b_2+\cdots+b_K}),\\
    \widetilde{\mathrm{SINR}}_{C} & = \frac{1}{K}(\frac{a_1}{b_1}+\frac{a_2}{b_2}+\cdots+\frac{a_K}{b_K}),
\end{flalign}
\end{subequations}
where $a_1,a_2,\cdots,a_K\!\geq\!0$ and $b_1,b_2,\cdots,b_K\!>\!0$ with $K\geq{1}$.
Based on Sedrakyan's inequality, we have 
\begin{flalign} \label{Sedrakyan}
  \frac{(\sqrt{a_1}+\sqrt{a_2}+\cdots+\sqrt{a_K})^2}{b_1+b_2+\cdots+b_K} \leq \frac{a_1}{b_1}+\frac{a_2}{b_2}+\cdots+\frac{a_K}{b_K}.
\end{flalign}
Further, leveraging on Cauchy-Schwarz inequality, we have  
\begin{flalign} \label{Cauchy-Schwarz}
  a_1+a_2+\cdots+a_K \leq (\sqrt{a_1}+\sqrt{a_2}+\cdots+\sqrt{a_K})^2.
\end{flalign}
Substituting \eqref{Sedrakyan} and \eqref{Cauchy-Schwarz} into \eqref{SINR_pro}, we can conclude that ${\mathrm{SINR}}_{C} \leq \widetilde{\mathrm{SINR}}_{C}$ and the equality holds if and only if $a_1=a_2=\cdots=a_K=0$. This completes the proof. 
\end{IEEEproof}
\subsection{Radar Receiver}
\label{ssec:Radar}
The  signal at the radar receiver follows a model similar to that of communications with the addition of the two way propagation. Similar to Section \ref{ssec:Comm_Mod}, after sampling, CP removal and applying $N_B$ $K$-point FFT, the echo signal on the $k$-th subcarrier at the radar receiver is
\begin{flalign}\label{radar_rec}
{\bf y}_{R}[f_k] = {{\sum}_{t=1}^{T}}{\bf H}_{R_t,k}{\bf M}{\bf V}_k{\bf F}_k{\bf s}_k
&+{{\sum}_{q=1}^{Q}}{\bf H}_{R_q,k}{\bf M}{\bf V}_k{\bf F}_k{\bf s}_k\!+\!{\bf n}_{R}[f_k],
\end{flalign} 
where 
the composite channels across DFBS-$t$-th target-receiver and 
DFBS-$q$-th clutter-receiver are, respectively, 
\begin{align}\label{radar_channel}
{\bf H}_{R_t,k} & = {\alpha_t}({\bf h}_{R_t,k}+{\bf G}_k^T{\bf\Phi}{\bf b}_{R_t,k})({\bf h}^T_{R_t,k}+{\bf b}_{R_t,k}^T{\bf\Phi}{\bf G}_k), \\
{\bf H}_{R_q,k} & = {\alpha_q}({\bf h}_{R_q,k}+{\bf G}_k^T{\bf\Phi}{\bf b}_{R_q,k})({\bf h}^T_{R_q,k}+{\bf b}_{R_q,k}^T{\bf\Phi}{\bf G}_k).
\end{align}
where ${\alpha_t}$ and ${\alpha_q}$ denote the RCS for the $t$-th target and $q$-th clutter, respectively, ${\bf h}_{R_{t},k}=g_{Bt,k}{\bf a}_B(f_k,\theta_{Bt},\psi_{Bt})$ denotes the path from DFBS to target and ${\bf b}_{R_t,k}=g_{Rt,k}{\bf a}_R(f_k,\theta_{Rt},\psi_{Rt})$ denotes the path from RIS to target.  

For the radar system, the performance of target detection is largely determined by the output SINR\footnote{Here, interference includes clutter response in addition to response from other targets.} and the detection performance for a given false-alarm improves with SINR. Thus, the maximization of SINR is widely used as the optimization criterion \cite{wu2017transmit,tsinos2021joint}. As in the communications system, in case of  multiple targets, the SINR for each of the targets need to be improved, particularly, improving the echo signal while suppressing interference. In this context, 
based on \eqref{radar_rec}, we define the radar SINR for the $t$-th target as 
\begin{align}\label{SINR_t}
   \footnotesize
   \mathrm{SINR}_{R_t} 
    &\!=\! \frac{{\sum_{k=1}^{K}}\|{\bf w}_{k,t}^H{\bf H}_{R_t,k}{\bf M}{\bf V}_k{\bf F}_k\|_2^2}{\sum_{j\neq t}^{T}\sum_{q=1}^{Q}\sum_{k=1}^{K}\|{\bf w}_{k,t}^H{\bf H}_{R_{j,q},k}{\bf M}{\bf V}_k{\bf F}_k\|_2^2\!+\!\sum_{k=1}^{K}\sigma_r^2{\bf w}_{k,t}^H{\bf w}_{k,t}},
\end{align} 
where ${\bf w}_{k,t}$ denotes the radar receive filter at the $k$-subcarrier \cite{cheng2018spectrally}.

\begin{remark}
Transmit beampattern matching is another approach widely employed as the optimal design criterion for radar sensing in the conventional and RHS-aided DFRC system without RIS \cite{cheng2021hb,liu2020joint,cheng2021hybrid,zhang2022holographic}. However, this may not be directly applicable for RIS-assisted radar-only or DFRC system, especially, in dense environments \cite{buzzi2022foundations,song2022joint} because it is difficult to focus the beam towards the target and RIS directions without pathloss information. For example, if the LoS path is totally blocked, then the allocation of power to the direct link is not needed. 
Further, in such environments with a weak or no line-of-sight channel, aggregated interference caused by multiple reflections from other objects in the environment could result in performance loss, despite the beampattern design.  Hence, the output SINR, which includes these artefacts and the design of received filter, is recognized as the proper radar metric for RIS-aided DFRC \cite{liu2022joint,hua2022joint}.
\end{remark}
%
%
\subsection{Problem Formulation}
Our goal in a radar-centric DFRC is to maximize the worst-case radar SINR while guaranteeing the communications SINR over all users. We formulate this optimization problem as
\begin{subequations} \label{opt_problem}
  \begin{align}    \mathop{{\mathrm{maximize}}}\limits_{{\bf w}_{k,t},{\bf w}_{k,u},{\bf F}_k,{\bf\Phi},{\bf M}} & {\quad} \min_{t}\textrm{SINR}_{R_t}\\
    \textrm{subject to} 
        & {\quad} |{\bf\Phi}|=1 \\
        & {\quad} \textrm{SINR}_{C_u}\geq \eta, \forall u,\\
        & {\quad}  0\leq{m}_{n_x^B,n_y^B}\leq1,  \\
        & {\quad} \|{\bf M}{\bf V}_k{\bf F}_k\|_F^2 \leq \mathcal{P}_k, \forall k,    
  \end{align}
\end{subequations}
where  $|{\bf\Phi}|=1$  indicates unit magnitude for each diagonal entry of the phase matrix ${\bf \Phi}$, $m_{n_x,n_y}\!=\!0$ denotes the $(n_x,n_y)$-th RHS element is disabled and $m_{n_x,n_y}\!=\!1$ denotes the $(n_x,n_y)$-th RHS element is unit gain, and $\eta$ denotes the threshold of communications user. Note that the above optimization problem involves the maximin objective function, difference of convex (DC). and unimodular constraints. Meanwhile, it is a fractional quadratically constrained quartic program (QCQP) problem in multiple variables and thus difficult to  solve directly. Despite the existence of several approaches for non-linear optimization addressing fractional QCQP for single variable, the problem in \eqref{opt_problem} poses unique challenges that prevent an adaptation of using existing methods. These challenges include, highly coupled variables,  maximin objective and several nonconvex constraints as well as presence of discrete variables. As a consequence, we develop the AO algorithm in the sequel. 
\section{Alternating Optimization}
\label{sec:typestyle}
%
We first decouple the nonconvex fractional QCQP into four subproblems of designing the receive filter along with digital, holographic, and passive beamformers. Then, we resort to AO procedure to solve these problems.  
\subsection{Sub-problem 1: Update of receive filter ${\bf w}_{k,t}$ and ${\bf w}_{k,u}$} 
We first define the communications filter ${\bf w}_u=[{\bf w}^T_{1,u},\cdots,{\bf w}^T_{K,u}]^T$ and radar filter ${\bf w}_t=[{\bf w}^T_{1,t},\cdots,{\bf w}^T_{K,t}]^T$ for all subcarriers. Then, for fixed ${\bf\Phi},{\bf F}_k$ and ${\bf M}$, the subproblem with respect to ${\bf w}_{k,t}$ and ${\bf w}_{k,u}$ is 
\begin{equation}\label{subproblem1}
\mathcal{P}_1\left\{
\begin{aligned}
    \mathop{{\mathrm{maximize}}}\limits_{{\bf w}_{t},{\bf w}_{u}} & {\quad} \min_{t}\frac{{\bf w}_t^H{\bf\Sigma}_{t}^{\mathcal{P}_1}{\bf w}_t}{{\bf w}_t^H\widetilde{\bf\Sigma}_{t}^{\mathcal{P}_1}{\bf w}_t+\sigma_r^2{\bf w}_t^H{\bf w}_t},  \\
    \textrm{subject to}       & {\quad} \frac{{\bf w}_u^H{\bf\Sigma}_{u}^{\mathcal{P}_1}{\bf w}_u}{{\bf w}_u^H\widetilde{\bf\Sigma}_{u}^{\mathcal{P}_1}{\bf w}_u+\sigma_c^2{\bf w}_u^H{\bf w}_u}\geq \hat{\eta}, \forall u
\end{aligned}
\right.
\end{equation}
where $\hat{\eta}=K\eta$ and the block diagonal matrix used in $\mathcal{P}_1$ is given by
\begin{subequations} \label{block_P1}
\begin{align}
{\bf\Sigma}_{t}^{\mathcal{P}_1} & = \mathrm{blkdiag}[{\bf\Sigma}_{1,t}^{\mathcal{P}_1},\cdots,{\bf\Sigma}_{K,t}^{\mathcal{P}_1}],
~\widetilde{\bf\Sigma}_{t}^{\mathcal{P}_1} = \mathrm{blkdiag}[\widetilde{\bf\Sigma}_{1,t}^{\mathcal{P}_1},\cdots,\widetilde{\bf\Sigma}_{K,t}^{\mathcal{P}_1}], ~t=1,\cdots,T, \\
{\bf\Sigma}_{u}^{\mathcal{P}_1} & = \mathrm{blkdiag}[{\bf\Sigma}_{1,u}^{\mathcal{P}_1},\cdots,{\bf\Sigma}_{K,u}^{\mathcal{P}_1}],
~\widetilde{\bf\Sigma}_{u}^{\mathcal{P}_1}  = \mathrm{blkdiag}[\widetilde{\bf\Sigma}_{1,u}^{\mathcal{P}_1},\cdots,\widetilde{\bf\Sigma}_{K,u}^{\mathcal{P}_1}], ~u=1,\cdots,U.
\end{align}
\end{subequations}
and 
\begin{subequations} \label{summation_P1}
\begin{align}
{\bf\Sigma}_{k,u}^{\mathcal{P}_1} & = {\bf H}_{C_u,k}{\bf M}{\bf V}_k{\bf F}_{k}{\bf\Lambda}{\bf F}_{k}^H{\bf V}_k^H{\bf M}^H{\bf H}_{C_u,k}^H,  
~\widetilde{\bf\Sigma}_{k,u}^{\mathcal{P}_1} ={\bf H}_{C_u,k}{\bf M}{\bf V}_k{\bf F}_{k}\widetilde{\bf\Lambda}{\bf F}_{k}^H{\bf V}_k^H{\bf M}^H{\bf H}_{C_u,k}^H,\\
{\bf\Sigma}_{k,t}^{\mathcal{P}_1} & = {\bf H}_{R_t,k}{\bf M}{\bf V}_k{\bf F}_{k}{\bf F}^H_{k}{\bf V}^H_k{\bf M}{\bf H}^H_{R_t,k},
~\widetilde{\bf\Sigma}_{k,t}^{\mathcal{P}_1} = \sum_{j\neq t}^{T}\sum_{q=1}^{Q}{\bf H}_{R_{j,q},k}{\bf M}{\bf V}_k{\bf F}_{k}{\bf F}^H_{k}{\bf V}^H_k{\bf M}{\bf H}^H_{R_{j,q},k}.
\end{align}
\end{subequations}
Note that the objective function of problem \eqref{subproblem1} is separable in terms of the variables ${\bf w}_t$ and ${\bf w}_u$. Hence, we obtain an optimal
solution for the maximin problem \eqref{subproblem1} by solving the following
disjoint problems
\begin{equation}\label{subproblem1.1}
 \mathcal{P}_{1.1}\left\{ 
  \begin{aligned}
    \mathop{{\mathrm{maximize}}}\limits_{{\bf w}_{t}} & {\quad} \frac{{\bf w}_t^H{\bf\Sigma}_{t}^{\mathcal{P}_1}{\bf w}_t}{{\bf w}_t^H\widetilde{\bf\Sigma}_{t}^{\mathcal{P}_1}{\bf w}_t+\sigma_r^2{\bf w}_t^H{\bf w}_t}, \\
    \mathop{{\mathrm{maximize}}}\limits_{{\bf w}_{u}} & {\quad} \frac{{\bf w}_u^H{\bf\Sigma}_{u}^{\mathcal{P}_1}{\bf w}_u}{{\bf w}_u^H\widetilde{\bf\Sigma}_{u}^{\mathcal{P}_1}{\bf w}_u+\sigma_c^2{\bf w}_u^H{\bf w}_u}.
  \end{aligned}
 \right.
\end{equation}
where $t=1,\cdots,T$ and $u=1,\cdots,U$. Note that $\mathcal{P}_{1.1}$ is composed by a set of generalized Rayleigh quotient programming which can be solved by the following Proposition~\ref{cor:opt}.

\begin{proposition} For a fixed Hermitian matrix ${\bf A}$, and a fixed positive definite matrix ${\bf B}$, the maximum values $\lambda^{\ast}$ of the generalized Rayleigh quotient $\frac{{\bf w}^H{\bf A}{\bf w}}{{\bf w}^H{\bf B}{\bf w}}$, where ${\bf w}\neq{\bf 0}$ and the corresponding vector ${\bf w}^{\ast}$ satisfy: 
$\lambda^{\ast} =\lambda_{\mathrm{max}}({\bf B}^{-1}{\bf A})$, ${\bf w}^{\ast} = \rho_{\mathrm{max}}({\bf B}^{-1}{\bf A})$,
where $\lambda_{\mathrm{max}}(\cdot)$ and $\rho_{\mathrm{max}}(\cdot)$ denotes the operation of largest eigenvalue and principal eigenvector. 
\label{cor:opt}
\end{proposition}
\begin{IEEEproof}
 See Appendix A.
\end{IEEEproof}
Based on Proposition~\ref{cor:opt}, the close-form solution of $\mathcal{P}_{1.1}$ is given by
\begin{subequations} \label{filter_opt}
  \begin{align}
    {\bf w}_t^{\ast} &= \rho_{\mathrm{max}}((\widetilde{\bf\Sigma}_{t}^{\mathcal{P}_1}+\sigma_r^2{\bf I}_{N_BK})^{-1}{{\bf\Sigma}_{t}^{\mathcal{P}_1}}), t=1,\cdots,T, \label{filter_opt_t}\\
    {\bf w}_u^{\ast} &= \rho_{\mathrm{max}}((\widetilde{\bf\Sigma}_{u}^{\mathcal{P}_1}+\sigma_c^2{\bf I}_{N_UK})^{-1}{{\bf\Sigma}_{u}^{\mathcal{P}_1}}), u=1,\cdots,U. \label{filter_opt_u}
  \end{align}
\end{subequations}
\subsection{Sub-problem 2: Update of digital beamforming ${\bf F}_k$}
For the fixed ${\bf w}_{k,t}$, ${\bf w}_{k,u}$, ${\bf M}$, and  ${\bf\Phi}$ , the subproblem with respect to ${\bf F}_{k}$ is  
\begin{equation}\label{subproblem2}
\mathcal{P}_{2}\left\{
\begin{aligned}
    \mathop{{\mathrm{maximize}}}\limits_{{\bf f}} & {\quad}\min_{t}\frac{{\bf f}^H{\bf\Sigma}_{t}^{\mathcal{P}_2}{\bf f}}{{\bf f}^H\widetilde{\bf\Sigma}_{t}^{\mathcal{P}_2}{\bf f}+\sigma_r^2{\bf w}_t^H{\bf w}_t} \\
    \textrm{subject to}      & {\quad} \|{\bf S}_k{\bf\Xi}{\bf f}\|_2^2 \leq \mathcal{P}_k, \forall k,\\
    & {\quad}  \frac{{\bf f}^H{\bf\Sigma}_{u}^{\mathcal{P}_2}{\bf f}}{{\bf f}^H\widetilde{\bf\Sigma}_{u}^{\mathcal{P}_2}{\bf f}+\sigma_c^2{\bf w}_u^H{\bf w}_u}\geq \hat{\eta}, \forall u,
\end{aligned}
\right.
\end{equation}
where ${\bf f}=[\mathrm{vec}({\bf F}_1)^T,\cdots,\mathrm{vec}({\bf F}_K)^T]^T$, ${\bf S}_k$ denotes the selection matrix to extract $k$-th interval of vector, ${\bf\Xi}=\mathrm{blkdiag}[({\bf I}_U\otimes{\bf M}{\bf V}_1),\cdots,({\bf I}_U\otimes{\bf M}{\bf V}_K)]$, and the block diagonal matrices ${\bf\Sigma}_{t}^{\mathcal{P}_2}$, $\widetilde{\bf\Sigma}_{t}^{\mathcal{P}_2}$, ${\bf\Sigma}_{u}^{\mathcal{P}_2}$ and $\widetilde{\bf\Sigma}_{u}^{\mathcal{P}_2}$ are, respectively, defined as
\begin{subequations} \label{block_P2}
\begin{align}
{\bf\Sigma}_{t}^{\mathcal{P}_2} & = \mathrm{blkdiag}[{\bf\Sigma}_{1,t}^{\mathcal{P}_2},\cdots,{\bf\Sigma}_{K,t}^{\mathcal{P}_2}],~  
\widetilde{\bf\Sigma}_{t}^{\mathcal{P}_2}  = \mathrm{blkdiag}[\widetilde{\bf\Sigma}_{1,t}^{\mathcal{P}_2},\cdots,\widetilde{\bf\Sigma}_{K,t}^{\mathcal{P}_2}], ~t=1,\cdots,T, \\
{\bf\Sigma}_{u}^{\mathcal{P}_2} & = \mathrm{blkdiag}[{\bf\Sigma}_{1,u}^{\mathcal{P}_2},\cdots,{\bf\Sigma}_{K,u}^{\mathcal{P}_2}], ~ 
\widetilde{\bf\Sigma}_{u}^{\mathcal{P}_2}  = \mathrm{blkdiag}[\widetilde{\bf\Sigma}_{1,u}^{\mathcal{P}_2},\cdots,\widetilde{\bf\Sigma}_{K,u}^{\mathcal{P}_2}], ~u=1,\cdots,U,
\end{align}
\end{subequations}
and
\begin{subequations} \label{summation_P2}
\begin{align}
{\bf\Sigma}_{k,u}^{\mathcal{P}_2} & = {\bf\Lambda}_u\otimes{{\bf V}_k^H{\bf M}^H{\bf H}_{C_u,k}^H{\bf w}_{k,u}{\bf w}_{k,u}^H{\bf H}_{C_u,k}{\bf M}{\bf V}_k}, \\
\widetilde{\bf\Sigma}_{k,u}^{\mathcal{P}_2} & =\widetilde{\bf\Lambda}_u\otimes{{\bf V}_k^H{\bf M}^H{\bf H}_{C_u,k}^H{\bf w}_{k,u}{\bf w}_{k,u}^H{\bf H}_{C_u,k}{\bf M}{\bf V}_k},\\
{\bf\Sigma}_{k,t}^{\mathcal{P}_2} & = {\bf I}_U\otimes{{\bf V}_k^H{\bf M}^H{\bf H}_{R_t,k}^H{\bf w}_{k,t}{\bf w}_{k,t}^H{\bf H}_{R_t,k}{\bf M}{\bf V}_k},  \\
\widetilde{\bf\Sigma}_{k,t}^{\mathcal{P}_2} & = {\bf I}_U\!\otimes\!\sum_{j\neq t}^{T}\sum_{q=1}^{Q}{{\bf V}_k^H{\bf M}^H{\bf H}_{R_{j,q},k}^H{\bf w}_{k,t}{\bf w}_{k,t}^H{\bf H}_{R_{j,q},k}{\bf M}{\bf V}_k}.
\end{align}
\end{subequations}
It is worth noting that ${\mathcal{P}_2}$ is highly nonconvex due to the fractional quadratic objective function and difference of convex (DC) constraint. Inspired by the minorization-maximization (MM) algorithm \cite{shen2019optimization}, we can linearize the corresponding convex function. 
Specifically, for the function $f({\bf x})={\bf x}^H{\bf H}{\bf x}$, the following inequality is always satisfied
\begin{equation} \label{inequality}
  f({\bf x})\geq 2\Re({{\bf x}^{(l)}}^H{\bf H}{\bf x})-f({{\bf x}^{(l)}}),
\end{equation}\normalsize
where ${\bf H}$ is {positive semidefinite (PSD)} matrix, ${{\bf x}^{(l)}}$ denotes the current point (at the $l$-th iteration), and the equality holds if and only if ${\bf x}={{\bf x}^{(l)}}$; See \cite{wei2022joint}.

Based on \eqref{inequality}, we simplify $\mathcal{P}_{2}$ as 
\begin{equation}\label{subproblem2.1}
\mathcal{P}_{2.1}\left\{
\begin{aligned}
    \mathop{{\mathrm{maximize}}}\limits_{{\bf f}} & ~ \min_{t}\frac{2\Re({{\bf f}^{(l)}}^H{\bf\Sigma}_{t}^{\mathcal{P}_2}{\bf f})\!-\!
    {{\bf f}^{(l)}}^H{\bf\Sigma}_{t}^{\mathcal{P}_2}{{\bf f}^{(l)}}}{{\bf f}^H\widetilde{\bf\Sigma}_{t}^{\mathcal{P}_2}{\bf f}+\sigma_r^2{\bf w}_t^H{\bf w}_t} \\
    \textrm{subject to}  & ~ \|{\bf S}_k{\bf\Xi}{\bf f}\|_2^2 \leq \mathcal{P}_k, \forall k, \\
    & ~  \frac{2\Re({{\bf f}^{(l)}}^H{\bf\Sigma}_{u}^{\mathcal{P}_2}{\bf f})\!-\!{{\bf f}^{(l)}}^H{\bf\Sigma}_{u}^{\mathcal{P}_2}{{\bf f}^{(l)}}}{{\bf f}^H\widetilde{\bf\Sigma}_{u}^{\mathcal{P}_2}{\bf f}+\sigma_c^2{\bf w}_u^H{\bf w}_u}\!\geq\! \hat{\eta}, \forall u,
\end{aligned}
\right.
\end{equation}
where ${\bf f}^{(l)}$ denotes the value of ${\bf f}$ at $l$-th outer AM iteration.  
This is a standard fractional maximin problem that can be solved using the generalized  Dinkelbach-based method  \cite{aubry2015optimizing}. Thus, we can solve problem \eqref{subproblem2.1} by reformulating it as
\begin{equation}\label{subproblem2.2}
\mathcal{P}_{2.1}\left\{
\begin{aligned}
    \mathop{{\mathrm{maximize}}}\limits_{{\bf f},\lambda_{\mathcal{P}_2}} & ~ \min_{t}{2\Re({{\bf f}^{(l)}}^H{\bf\Sigma}_{t}^{\mathcal{P}_2}{\bf f})}-\lambda_{\mathcal{P}_2}{\bf f}^H\widetilde{\bf\Sigma}_{t}^{\mathcal{P}_2}{\bf f} \\
    \textrm{subject to}  & ~ \|{\bf S}_k{\bf\Xi}{\bf f}\|_2^2 \leq \mathcal{P}_k, \forall k, \\
    & ~  \frac{2\Re({{\bf f}^{(l)}}^H{\bf\Sigma}_{u}^{\mathcal{P}_2}{\bf f})\!-\!{{\bf f}^{(l)}}^H{\bf\Sigma}_{u}^{\mathcal{P}_2}{{\bf f}^{(l)}}}{{\bf f}^H\widetilde{\bf\Sigma}_{u}^{\mathcal{P}_2}{\bf f}+\sigma_c^2{\bf w}_u^H{\bf w}_u}\!\geq\! \hat{\eta}, \forall u,
\end{aligned}
\right.
\end{equation}   
and solving using \textbf{Algorithm \ref{alg:DBS}}. Noted that for the simplified two variable quadratic programming with convex constraints \eqref{subproblem2.2}, Dinkelbach algorithm can be convergent to the global optimal solution \cite{aubry2015optimizing}.   

\begin{algorithm}[H]
	\caption{Dinkelbach-based algorithm to solve $\mathcal{P}_{2.1}$}
    \label{alg:DBS}
	\begin{algorithmic}[1]
		\Statex \textbf{Input:}  $\zeta_1$, ${\bf f}^{(l)}$, $\mathcal{P}_k$, ${\bf S}_k$, ${\bf\Xi}$, ${\bf\Sigma}_{t}^{\mathcal{P}_2}$, $\widetilde{\bf\Sigma}_{t}^{\mathcal{P}_2}$, ${\bf\Sigma}_{u}^{\mathcal{P}_2}$ and $\widetilde{\bf\Sigma}_{u}^{\mathcal{P}_2}$. \;\;
        \Statex 
        \textbf{Output:}
        ${\bf f}^{(l+1)}$
    
\vspace{-0.1cm}    
\State Set $l_2=0$, ${\bf f}_{l_2}={\bf f}^{(l)}$;
\vspace{-0.1cm}    
\State $\lambda_{\mathcal{P}_2}^{(l_2)}=\min_{t}\frac{2\Re({{\bf f}^{(l)}}^H{\bf\Sigma}_{t}^{\mathcal{P}_2}{\bf f})\!-\!
    {{\bf f}^{(l)}}^H{\bf\Sigma}_{t}^{\mathcal{P}_2}{{\bf f}^{(l)}}}{{\bf f}^H\widetilde{\bf\Sigma}_{t}^{\mathcal{P}_2}{\bf f}+\sigma_r^2{\bf w}_t^H{\bf w}_t}$
\vspace{-0.1cm}    
\Repeat
\vspace{-0.1cm}      
\State Find ${\bf f}_{l_2}$ by solving problem \eqref{subproblem2.2} using $\mathcal{P}_k$, ${\bf R}_{u}$, $\widetilde{\bf R}_{u}$, ${\bf\Xi}_{p,t}$, and ${\bf\Xi}_{p,c}$;
\vspace{-0.1cm}      
\State 
       $F_{\lambda_{\mathcal{P}_2}^{(l_2)}}=\min_{t} 2\Re({{\bf f}^{(l)}}^H{\bf\Sigma}_{t}^{\mathcal{P}_2}{\bf f}_{l_2})-\lambda_f^{(l_2)}{{\bf f}_{l_2}^H\widetilde{\bf\Sigma}_{t}^{\mathcal{P}_2}{\bf f}_{l_2}}$;
\vspace{-0.1cm}      
\State $l_2\leftarrow l_2+1$;
\vspace{-0.1cm}      
\State Update $\lambda_{\mathcal{P}_2}^{(l_2)}=\min_{t}\frac{2\Re({{\bf f}^{(l)}}^H{\bf\Sigma}_{t}^{\mathcal{P}_2}{\bf f}_{l_2})\!-\!
    {{\bf f}^{(l)}}^H{\bf\Sigma}_{t}^{\mathcal{P}_2}{{\bf f}^{(l)}}}{{\bf f}_{l_2}^H\widetilde{\bf\Sigma}_{t}^{\mathcal{P}_2}{\bf f}_{l_2}+\sigma_r^2{\bf w}_t^H{\bf w}_t}$;
\vspace{-0.1cm}    
\Until{$F_{\lambda_{l_2}}\leq \zeta_1$ or reach the maximum iteration. }
\vspace{-0.1cm}    
\State \Return ${\bf f}^{(l+1)}={\bf f}_{l_2}$.
  \end{algorithmic}
\end{algorithm}

\subsection{Sub-problem 3: Update of holographic beamforming ${\bf M}$}
With  ${\bf w}_{k,t}$, ${\bf w}_{k,u}$, ${\bf F}_k$, and  ${\bf\Phi}$ fixed, the subproblem with respect to ${\bf M}$ is
\begin{equation}\label{subproblem3}
\mathcal{P}_{3}\left\{
\begin{aligned}
    \mathop{{\mathrm{maximize}}}\limits_{{\bf 0}\preceq{\bf m}\preceq{\bf 1}} & {\quad} \min_{t}\frac{{\bf m}^T\Re({\bf\Sigma}_{t}^{\mathcal{P}_3}){\bf m}}{{\bf m}^T\Re(\widetilde{\bf\Sigma}_{t}^{\mathcal{P}_3}){\bf m}+\sigma_r^2{\bf w}_t^H{\bf w}_t}\\
    \textrm{subject to}  & {\quad} \|\mathrm{diag}({\bf m}){\bf V}_k{\bf F}_k\|_F^2 \leq \mathcal{P}_k, \forall k,\\
     & {\quad}  \frac{{\bf m}^T\Re({\bf\Sigma}_{u}^{\mathcal{P}_3}){\bf m}}{{\bf m}^T\Re(\widetilde{\bf\Sigma}_{u}^{\mathcal{P}_3}){\bf m}+\sigma_c^2{\bf w}_u^H{\bf w}_u}\geq \hat{\eta}, \forall u,\\
\end{aligned}
\right.
\end{equation}
where ${\bf m}={\bf M}^T{\bf 1}_{N_xN_y}$ and 
the matrices ${\bf\Sigma}_{t}^{\mathcal{P}_3}$, $\widetilde{\bf\Sigma}_{t}^{\mathcal{P}_3}$, ${\bf\Sigma}_{u}^{\mathcal{P}_3}$ and $\widetilde{\bf\Sigma}_{u}^{\mathcal{P}_3}$ are similarly defined as
\begin{subequations} \label{summation_P3}
\begin{align}
&{\bf\Sigma}_{u}^{\mathcal{P}_3}  = {\sum_{k=1}^{K}}\langle{\bf w}_{k,u}^H{\bf H}_{C_u,k}\rangle{\bf V}_k{\bf F}_k{\bf\Lambda}_u{\bf F}_k^H{\bf V}_k^H\langle{\bf w}_{k,u}^T{\bf H}^{\ast}_{C_u,k}\rangle, \\
&\widetilde{\bf\Sigma}_{u}^{\mathcal{P}_3} = {\sum_{k=1}^{K}}\langle{\bf w}_{k,u}^H{\bf H}_{C_u,k}\rangle{\bf V}_k{\bf F}_k\widetilde{\bf\Lambda}_u{\bf F}_k^H{\bf V}_k^H\langle{\bf w}_{k,u}^T{\bf H}^{\ast}_{C_u,k}\rangle,\\
&{\bf\Sigma}_{t}^{\mathcal{P}_3}  = {\sum_{k=1}^{K}}\langle{\bf w}_{k,t}^H{\bf H}_{R_t,k}\rangle{\bf V}_k{\bf F}_k{\bf F}_k^H{\bf V}_k^H\langle{\bf w}_{k,t}^T{\bf H}^{\ast}_{R_t,k}\rangle,  \\
 &\widetilde{\bf\Sigma}_{t}^{\mathcal{P}_3} = \sum_{j\neq t}^{T}\sum_{q=1}^{Q}{\sum_{k=1}^{K}}\langle{\bf w}_{k,t}^H{\bf H}_{R_{j,q},k}\rangle{\bf V}_k{\bf F}_k{\bf F}_k^H{\bf V}_k^H\langle{\bf w}_{k,t}^T{\bf H}^{\ast}_{R_{j,q},k}\rangle.
\end{align}
\end{subequations}

Similar to $\mathcal{P}_{2}$, we reformulate $\mathcal{P}_{3}$ as 
\begin{equation}\label{subproblem3.1}
\mathcal{P}_{3.1}\left\{
\begin{aligned}
    \mathop{{\mathrm{maximize}}}\limits_{{\bf 0}\preceq{\bf m}\preceq{\bf 1}} &~ \min_{t}\frac{2{{\bf m}^{(l)}}^T\Re({\bf\Sigma}_{t}^{\mathcal{P}_3}){\bf m}\!-\!{{\bf m}^{(l)}}^T\Re({\bf\Sigma}_{t}^{\mathcal{P}_3}){{\bf m}^{(l)}}}{{\bf m}^T\Re(\widetilde{\bf\Sigma}_{t}^{\mathcal{P}_3}){\bf m}+\sigma_r^2{\bf w}_t^H{\bf w}_t}\\
    \textrm{subject to}  & ~ \|\mathrm{diag}({\bf m}){\bf V}_k{\bf F}_k\|_F^2 \leq \mathcal{P}_k, \forall k,\\
     & ~  \frac{2{{\bf m}^{(l)}}^T\Re({\bf\Sigma}_{u}^{\mathcal{P}_3}){\bf m}\!-\!{{\bf m}^{(l)}}^T\Re({\bf\Sigma}_{u}^{\mathcal{P}_3}){{\bf m}^{(l)}}}{{\bf m}^T\Re(\widetilde{\bf\Sigma}_{u}^{\mathcal{P}_3}){\bf m}+\sigma_c^2{\bf w}_u^H{\bf w}_u}\!\geq\! \hat{\eta}, \forall u.\\
\end{aligned}
\right.
\end{equation}
This subproblem is similar to the previous $\mathcal{P}_{2.1}$ and  hence it can be also solved by \textbf{Algorithm \ref{alg:DBS}} with variables appropriately substituted.  Notice that, based on the inequality \eqref{inequality}, the objective value in \eqref{subproblem3} is always equal or great than the simplified problem \eqref{subproblem3.1} which guarantees the monotonic increasing of radar SINR in the MM iteration.  
\subsection{Sub-problem 4: Update of passive beamforming ${\bf\Phi}$}
With  ${\bf w}_{k,t}$, ${\bf w}_{k,u}$, ${\bf F}_k$, ${\bf M}$ fixed, the subproblem with respect to phase-shift design is
\begin{equation}\label{subproblem4}
\mathcal{P}_{4}\left\{
\begin{aligned}   \mathop{{\mathrm{maximize}}}\limits_{{\bm\phi},{\bm\varphi}} & ~\min_{t}\textrm{SINR}_{R_t} \\
    \textrm{subject to}  & ~ |{\bm\phi}|=1, |{\bm\varphi}|=1, {\bm\phi}={\bm\varphi}, \\
                         & ~ \frac{{\bm\phi}^T{\bf\Sigma}_{u}^{\mathcal{P}_4}{\bm\phi}^{\ast}\!+\!2\Re({\bm\phi}^T{\bf d}_u^{\mathcal{P}_4})\!+\!d_u^{\mathcal{P}_4}}{{\bm\phi}^T\widetilde{\bf\Sigma}_{u}^{\mathcal{P}_4}{\bm\phi}^{\ast}\!+\!2\Re({\bm\phi}^T\widetilde{\bf d}_u^{\mathcal{P}_4})\!+\!\tilde{d}_u^{\mathcal{P}_4}}\!\geq\!\hat{\eta}, \forall u,
\end{aligned}
\right.
\end{equation}
where ${\bm\phi}={\bf\Phi}^T{\bf 1}_{N_R}$ denotes the  phase-shift vector, ${\bm\varphi}$ denotes the auxiliary variable, $d_u^{\mathcal{P}_4}={\sum_{k=1}^{K}}d_{k,u}^{\mathcal{P}_4}$, ${\bf d}_u^{\mathcal{P}_4}=\sum_{k=1}^{K}{\bf d}_{k,u}^{\mathcal{P}_4}$, ${\bf\Sigma}_{u}^{\mathcal{P}_4}=\sum_{k=1}^{K}{\bf\Sigma}_{k,u}^{\mathcal{P}_4}$,
$\tilde{d}_u^{\mathcal{P}_4}={\sum_{k=1}^{K}}\tilde{d}_{k,u}^{\mathcal{P}_4}+\sigma_c^2{\bf w}_u^H{\bf w}_u$, $\widetilde{\bf d}_u^{\mathcal{P}_4}=\sum_{k=1}^{K}\widetilde{\bf d}_{k,u}^{\mathcal{P}_4}$, $\widetilde{\bf\Sigma}_{u}^{\mathcal{P}_4}=\sum_{k=1}^{K}\widetilde{\bf\Sigma}_{k,u}^{\mathcal{P}_4}$, and $d_{k,u}^{\mathcal{P}_4}$, ${\bf d}_{k,u}^{\mathcal{P}_4}$, ${\bf\Sigma}_{k,u}^{\mathcal{P}_4}$, $\tilde{d}_{k,u}^{\mathcal{P}_4}$, $\widetilde{\bf d}_{k,u}^{\mathcal{P}_4}$ and $\widetilde{\bf\Sigma}_{k,u}^{\mathcal{P}_4}$ are defined as following
\begin{subequations} \label{summation_P4}
\begin{align}
\hspace*{-0.1in} d_{k,u}^{\mathcal{P}_4} & \!=\! {\bf w}_{k,u}^H{\bf H}^{\mathrm{dir}}_{C_u,k}{\bf M}{\bf V}_k{\bf F}_k 
{\bf\Lambda}_u({\bf w}_{k,u}^H{\bf H}^{\mathrm{dir}}_{C_u,k}{\bf M}{\bf V}_k{\bf F}_k)^H, \\
\hspace*{-0.1in} {\bf d}_{k,u}^{\mathcal{P}_4} & = \langle{\bf w}_{k,u}^H{\bf H}^{\mathrm{RIS}}_{C_u,k}\rangle{\bf G}_k{\bf M}{\bf V}_k{\bf F}_k{\bf\Lambda}_u({\bf w}_{k,u}^H{\bf H}^{\mathrm{dir}}_{C_u,k}{\bf M}{\bf V}_k{\bf F}_k)^H,\\
\hspace*{-0.1in} {\bf\Sigma}_{k,u}^{\mathcal{P}_4} & = \langle{\bf w}_{k,u}^H{\bf H}^{\mathrm{RIS}}_{C_u,k}\rangle{\bf G}_k{\bf M}{\bf V}_k{\bf F}_k{\bf\Lambda}_u(\langle{\bf w}_{k,u}^H{\bf H}^{\mathrm{RIS}}_{C_u,k}\rangle{\bf G}_k{\bf M}{\bf V}_k{\bf F}_k)^H, \\
\hspace*{-0.1in} \tilde{d}_{k,u}^{\mathcal{P}_4} & \!=\! {\bf w}_{k,u}^H{\bf H}^{\mathrm{dir}}_{C_u,k}{\bf M}{\bf V}_k{\bf F}_k 
\widetilde{\bf\Lambda}_u({\bf w}_{k,u}^H{\bf H}^{\mathrm{dir}}_{C_u,k}{\bf M}{\bf V}_k{\bf F}_k)^H, \\
\hspace*{-0.1in} \widetilde{\bf d}_{k,u}^{\mathcal{P}_4} & = \langle{\bf w}_{k,u}^H{\bf H}^{\mathrm{RIS}}_{C_u,k}\rangle{\bf G}_k{\bf M}{\bf V}_k{\bf F}_k\widetilde{\bf\Lambda}_u({\bf w}_{k,u}^H{\bf H}^{\mathrm{dir}}_{C_u,k}{\bf M}{\bf V}_k{\bf F}_k)^H,\\
\hspace*{-0.1in} \widetilde{\bf\Sigma}_{k,u}^{\mathcal{P}_4} & = \langle{\bf w}_{k,u}^H{\bf H}^{\mathrm{RIS}}_{C_u,k}\rangle{\bf G}_k{\bf M}{\bf V}_k{\bf F}_k\widetilde{\bf\Lambda}_u(\langle{\bf w}_{k,u}^H{\bf H}^{\mathrm{RIS}}_{C_u,k}\rangle{\bf G}_k{\bf M}{\bf V}_k{\bf F}_k)^H.
\end{align}
\end{subequations}
%
Accordingly, $\mathcal{P}_{4}$ is reformulated as 
\begin{equation}\label{subproblem4.1}
\mathcal{P}_{4.1}\left\{
\begin{aligned}
    \mathop{{\mathrm{maximize}}}\limits_{{\bm\phi},{\bm\varphi}} & ~ \min_{t}\frac{f_t({\bm\phi},{\bm\varphi})}{g_t({\bm\phi},{\bm\varphi})} \\
    \textrm{subject to}  & ~ |{\bm\phi}|=1, |{\bm\varphi}|=1, {\bm\phi}-{\bm\varphi}={\bf0}, \\
                         & ~ \frac{{\bm\phi}^T{\bf\Sigma}_{u}^{\mathcal{P}_4}{\bm\phi}^{\ast}\!+\!2\Re({\bm\phi}^T{\bf d}_u^{\mathcal{P}_4})\!+\!d_u^{\mathcal{P}_4}}{{\bm\phi}^T\widetilde{\bf\Sigma}_{u}^{\mathcal{P}_4}{\bm\phi}^{\ast}\!+\!2\Re({\bm\phi}^T\widetilde{\bf d}_u^{\mathcal{P}_4})\!+\!\tilde{d}_u^{\mathcal{P}_4}}\!\geq\!\hat{\eta}, \forall u,
\end{aligned}
\right.
\end{equation}
where 
\begin{subequations} \label{summation_P4.0}
\begin{align}
f_t({\bm\phi},{\bm\varphi}) & =  {\bm\phi}^T{\bf\Sigma}_{t}^{\mathcal{P}_4}({\bm\varphi}){\bm\phi}^{\ast}\!+\!2\Re({\bm\phi}^T{\bf d}_t^{\mathcal{P}_4}({\bm\varphi}))\!+\!d_t^{\mathcal{P}_4}({\bm\varphi}) \nonumber \\
& = {\bm\varphi}^T{\bf\Sigma}_{t}^{\mathcal{P}_4}({\bm\phi}){\bm\varphi}^{\ast}\!+\!2\Re({\bm\varphi}^T{\bf d}_t^{\mathcal{P}_4}({\bm\phi}))\!+\!d_t^{\mathcal{P}_4}({\bm\phi}), \label{summation_P4.0a}\\
g_t({\bm\phi},{\bm\varphi}) & =  {\bm\phi}^T\widetilde{\bf\Sigma}_{t}^{\mathcal{P}_4}({\bm\varphi}){\bm\phi}^{\ast}\!+\!2\Re({\bm\phi}^T\widetilde{\bf d}_t^{\mathcal{P}_4}({\bm\varphi}))\!+\!\tilde{d}_t^{\mathcal{P}_4}({\bm\varphi}) \nonumber\\
& = {\bm\varphi}^T\widetilde{\bf\Sigma}_{t}^{\mathcal{P}_4}({\bm\phi}){\bm\varphi}^{\ast}\!+\!2\Re({\bm\varphi}^T\widetilde{\bf d}_t^{\mathcal{P}_4}({\bm\phi}))\!+\!\tilde{d}_t^{\mathcal{P}_4}({\bm\phi}). \label{summation_P4.0b}
\end{align}
\end{subequations}
and ${\bf\Sigma}_{t}^{\mathcal{P}_4}({\bm\varphi})$, ${\bf d}_t^{\mathcal{P}_4}({\bm\varphi})$, $d_t^{\mathcal{P}_4}({\bm\varphi})$,  ${\bf\Sigma}_{t}^{\mathcal{P}_4}({\bm\phi})$, ${\bf d}_t^{\mathcal{P}_4}({\bm\phi})$, and $d_t^{\mathcal{P}_4}({\bm\phi})$ are defined as 
\begin{subequations}\label{summation_P4.1}
 \footnotesize
  \begin{align}	
d_t^{\mathcal{P}_4}({\bm\varphi}) & \!=\! {\sum_{k=1}^{K}}\|{\alpha_t}{\bf w}_{k,t}^H({\bf h}_{R_t,k}{\bf h}_{R_t,k}^T\!+\!{\bf h}_{R_t,k}{\bf b}_{R_t,k}^T\tilde{\bm\varphi}{\bf G}_k){\bf M}{\bf V}_k{\bf F}_k \|_2^2, \\
{\bf d}_t^{\mathcal{P}_4}({\bm\varphi}) & \!= \!{\sum_{k=1}^{K}}{\alpha_t}^2\langle{\bf w}_{k,t}^H{\bf G}^T_{k}\rangle({\bf b}_{R_t,k}{\bf h}_{R_t,k}^T\!+\!{\bf b}_{R_t,k}{\bf b}_{R_t,k}^T\tilde{\bm\varphi}{\bf G}_k){\bf M}{\bf V}_k{\bf F}_k{\bf F}_k^H{\bf V}_k^H{\bf M}({\bf h}_{R_t,k}{\bf h}_{R_t,k}^T\!+\!{\bf h}_{R_t,k}{\bf b}_{R_t,k}^T\tilde{\bm\varphi}{\bf G}_k)^H{\bf w}_{k,t},   \\
{\bf\Sigma}_{t}^{\mathcal{P}_4}({\bm\varphi}) & \!=\!  {\sum_{k=1}^{K}}{\alpha_t}^2\langle{\bf w}_{k,t}^H{\bf G}^T_{k}\rangle({\bf b}_{R_t,k}{\bf h}_{R_t,k}^T\!+\!{\bf b}_{R_t,k}{\bf b}_{R_t,k}^T\tilde{\bm\varphi}{\bf G}_k){\bf M}{\bf V}_k{\bf F}_k{\bf F}_k^H{\bf V}_k^H{\bf M}(\langle{\bf w}_{k,t}^H{\bf G}^T_{k}\rangle({\bf b}_{R_t,k}{\bf h}_{R_t,k}^T\!+\!{\bf b}_{R_t,k}{\bf b}_{R_t,k}^T\tilde{\bm\varphi}{\bf G}_k))^H, \\
d_t^{\mathcal{P}_4}({\bm\phi}) & \!=\! {\sum_{k=1}^{K}}\|{\alpha_t}{\bf w}_{k,t}^H({\bf h}_{R_t,k}{\bf h}_{R_t,k}^T\!+\!{\bf G}_k^T{\bf\Phi}{\bf b}_{R_t,k}{\bf h}_{R_t,k}^T){\bf M}{\bf V}_k{\bf F}_k \|_2^2,  \\
{\bf d}_t^{\mathcal{P}_4}({\bm\phi}) & \!=\!  {\sum_{k=1}^{K}}{\alpha_t}^2\langle{\bf w}_{k,t}^H({\bf h}_{R_t,k}+{\bf G}_k^T{\bf\Phi}{\bf b}_{R_t,k}){\bf b}_{R_t,k}\rangle{\bf G}_k{\bf M}{\bf V}_k{\bf F}_k{\bf F}_k^H{\bf V}_k^H{\bf M}({\bf h}_{R_t,k}{\bf h}_{R_t,k}+{\bf G}_k^T{\bf\Phi}{\bf b}_{R_t,k}{\bf h}_{R_t,k})^H{\bf w}_{k,t},   \\
{\bf\Sigma}_{t}^{\mathcal{P}_4}({\bm\phi}) & \!=\!   {\sum_{k=1}^{K}}{\alpha_t}^2\langle{\bf w}_{k,t}^H({\bf h}_{R_t,k}\!+\!{\bf G}_k^T{\bf\Phi}{\bf b}_{R_t,k}){\bf b}_{R_t,k}\rangle{\bf G}_k{\bf M}{\bf V}_k{\bf F}_k{\bf F}_k^H{\bf V}_k^H{\bf M}(\langle{\bf w}_{k,t}^H({\bf h}_{R_t,k}\!+\!{\bf G}_k^T{\bf\Phi}{\bf b}_{R_t,k}){\bf b}_{R_t,k}\rangle{\bf G}_k)^H.
  \end{align}
\end{subequations} 
where $\tilde{\bm\varphi}=\mathrm{diag}(\bm\varphi)$, and the variables related to \eqref{summation_P4.0b} can be similarly defined as \eqref{summation_P4.0a} and hence omit it herein.

Rewriting the problem $\mathcal{P}_{4.1}$ as
\begin{equation}\label{subproblem4.2}
\mathcal{P}_{4.2}\left\{
\begin{aligned}
    \mathop{{\mathrm{maximize}}}\limits_{{\bm\phi},{\bm\varphi}} & ~ \min_{t}f_t({\bm\phi},{\bm\varphi})-\lambda_{\mathcal{P}_4}g_t({\bm\phi},{\bm\varphi}) \\
    \textrm{subject to}  & ~ |{\bm\phi}|=1, |{\bm\varphi}|=1, {\bm\phi}-{\bm\varphi}={\bf0}, \\
                         & ~ \frac{{\bm\phi}^T{\bf\Sigma}_{u}^{\mathcal{P}_4}{\bm\phi}^{\ast}\!+\!2\Re({\bm\phi}^T{\bf d}_u^{\mathcal{P}_4})\!+\!d_u^{\mathcal{P}_4}}{{\bm\phi}^T\widetilde{\bf\Sigma}_{u}^{\mathcal{P}_4}{\bm\phi}^{\ast}\!+\!2\Re({\bm\phi}^T\widetilde{\bf d}_u^{\mathcal{P}_4})\!+\!\tilde{d}_u^{\mathcal{P}_4}}\!\geq\!\hat{\eta}, \forall u,
\end{aligned}
\right.
\end{equation}
where $\lambda_{\mathcal{P}_4}\!=\!\min_{t}\frac{f_t({\bm\phi},{\bm\varphi})}{g_t({\bm\phi},{\bm\varphi})}$ denotes the Dinkelbach parameter. According to \eqref{inequality}, we further simplify problem $\mathcal{P}_{4.2}$ as \cite{yang2020dual}  
\begin{equation}\label{subproblem4.3}
\mathcal{P}_{4.3}\left\{
\begin{aligned}
    \mathop{{\mathrm{minimize}}}\limits_{{\bm\phi},{\bm\varphi}} & ~ \max_{\|{\bf z}\|_1=1} \sum_{t=1}^{T}z_t(\lambda_{\mathcal{P}_4}g_t({\bm\phi},{\bm\varphi})-\hat{f}_t({\bm\phi},{\bm\varphi})) \\
    \textrm{subject to}  & ~ |{\bm\phi}|=1, |{\bm\varphi}|=1, {\bm\phi}-{\bm\varphi}={\bf0}, \\
      & ~ 2\Re({\bm\phi}^T{\bf p}_u^{\mathcal{P}_4})\leq p_u, \forall u.
\end{aligned}
\right.
\end{equation}
where ${\bf z}=[z_1,\cdots,z_T]$ denotes the weight vector, ${\hat f}_t({\bm\phi},{\bm\varphi})$ is the right side of inequality \eqref{inequality} and thus a lower bound function of ${ f}_t({\bm\phi},{\bm\varphi})$, and ${\bf p}_u^{\mathcal{P}_4}$ is the reconstructed vector to linearize the SINR constraint. 
Then, the augmented Lagrangian function of \eqref{subproblem4.3} is
\begin{align}\label{Lagrangian}
  \mathcal{L}({\bm\phi},{\bm\psi},{\bf u},{\bf w},{\rho})=
     \max_{\|{\bf z}\|_1=1} \sum_{t=1}^{T}z_t(\lambda_{\mathcal{P}_4}g_t({\bm\phi},{\bm\varphi})-\hat{f}_t({\bm\phi},{\bm\varphi})) +\frac{{\rho}}{2}\|{\bm\phi}-{\bm\psi}+{\bf u}\|_2^2 +\Re\{{\bf w}^T{\bf c}\},
\end{align}
where $\rho$ is the penalty parameter, ${\bf u}$ and ${\bf w}$ denote the auxiliary variables, and ${\bf c}=[c_1,\cdots,c_U]^T$, in which $c_u=2\Re({\bm\phi}^T{\bf p}_u^{\mathcal{P}_4})-p_u$.
Then, as the previous work problem \cite{yang2020dual}, $\mathcal{P}_{4.3}$ is solved by the C-ADMM algorithm  which is summarized in \textbf{Algorithm~\ref{alg:C-ADMM}}. Noted that in each ADMM iteration, the nonconvex  unit-sphere programming \eqref{ADMM_sub_phi} and \eqref{ADMM_sub_psi} can be directly solved by RSD algorithm, see details in \cite{alhujaili2019transmit}.

\begin{algorithm}[H]
  \caption{C-ADMM algorithm to solve $\mathcal{P}_{4.1}$}
  \label{alg:C-ADMM}
  \begin{algorithmic}[1]
    \Statex {\bf Input:} $\zeta_2$, ${\bf u}$, ${\bf w}$, and  ${{\bm\phi}^{(l)}}$ \;\;
    \Statex 
    {\bf Output:} ${{\bm\phi}^{(l+1)}}={\bm\phi}_{l_4}$.
\vspace{-0.15cm}    
\State Set $l_4=0$, ${{\bm\psi}^{(l)}}={{\bm\phi}^{(l)}}$;
\vspace{-0.15cm}    
\Repeat
\vspace{-0.15cm}      
\State Compute: ${\lambda}_{\mathcal{P}_4}^{(l_4)}=\min_{t}\frac{f_t({\bm\phi},{\bm\varphi})}{g_t({\bm\phi},{\bm\varphi})}$;
\vspace{-0.15cm} 
\State Update ${\bf z}$ by solving the problem \eqref{subproblem4.3}, which is convex w.r.t. the variable ${\bf z}$;
\State Update ${\bm\phi}_{l_4}$ via solving 
\vspace{-0.15cm} 
 \begin{equation}\label{ADMM_sub_phi}
           \mathop{{\mathrm{minimize}}}\limits_{{\bm\phi}}~\mathcal{L}({\bm\phi},{\bm\psi}^{(l_4)},{\bf u},{\bf w},{\rho}),\quad
               \textrm{subject to} ~ |{\bm\phi}|=1.
      \end{equation}
\State Update ${\bm\psi}_{l_4}$ via solving
  \begin{equation}\label{ADMM_sub_psi}
           \mathop{{\mathrm{minimize}}}\limits_{{\bm\psi}}~ \mathcal{L}({\bm\phi}^{(l_4+1)},{\bm\psi},{\bf u},{\bf w},{\rho}),\quad
            \textrm{subject to} ~ |{\bm\psi}|=1.
  \end{equation}
\State Update the dual variable ${\bf u}$ and ${\bf w}$;
\vspace{-0.15cm}     
\State $l_4\leftarrow l_4+1$;
\vspace{-0.15cm}    
\Until{$\|{\bm\phi}_{l_4}\!-\!{\bm\phi}_{l_4-1}\|_2^2\!\leq\!\zeta_2$  or reach the maximum iteration};
\vspace{-0.15cm}    
\State \Return ${\bm\phi}_{l_4}$;
  \end{algorithmic}
\end{algorithm}

Based on above, the proposed AM algorithm for the optimization problem \eqref{opt_problem} is summarized in \textbf{Algorithm~\ref{alg:AM}}.

\begin{algorithm}[H]
  \caption{Alternating maximization algorithm to solve \eqref{opt_problem}}
  \label{alg:AM}
  \begin{algorithmic}[1]
		\Statex \textbf{Input:} $\zeta_3$, ${\bf w}_t^{(l)}$, ${\bf w}_u^{(l)}$, ${\bf f}^{(l)}$, ${\bf M}^{(l)}$ and ${\bm\phi}^{(l)}$. \;\;
		\Statex 
		\textbf{Output:} ${\bf w}_t^{\star}={\bf w}^{(l)}_t$, ${\bf w}_u^{\star}={\bf w}^{(l)}_u$, ${\bf F}_k^{\star}={\bf F}_k^{(l)}$, ${\bf M}^{\star}={\bf M}^{(l)}$ and ${\bf\Phi}^{\star}={\bf\Phi}^{(l)}$. 
\vspace{-0.15cm}    
\State Set $l=0$;
\vspace{-0.15cm}    
\Repeat
\vspace{-0.15cm}      
\State Update ${\bf w}^{(l)}_t$ and ${\bf w}^{(l)}_u$ as \eqref{filter_opt_t} and \eqref{filter_opt_u}, respectively;
\vspace{-0.15cm}      
\State Update ${\bf f}^{(l)}$ via \textbf{Algorithm~\ref{alg:DBS}} and reconstruct ${\bf F}_k^{(l)}$;
\vspace{-0.15cm} 
\State Update ${\bf M}^{(l)}$ via \textbf{Algorithm~\ref{alg:DBS}} by changing the input variables;
\vspace{-0.15cm}    
\State Update ${\bm\phi}^{(l)}$ via \textbf{Algorithms~\ref{alg:C-ADMM}} and reconstruct ${\bf \Phi}^{(l)}$;
\vspace{-0.15cm}      
\State $l\leftarrow l+1$;
\vspace{-0.15cm}    
\Until{$(\mathrm{\min}_{t}\mathrm{SINR}_{R_t}^{(l)}-\mathrm{\min}_{t}\mathrm{SINR}_{R_t}^{(l-1)})^2\leq\zeta_3$ or maximum iteration reached}; 
\vspace{-0.15cm}    
\State \Return ${\bf w}^{(l)}_t$, ${\bf w}^{(l)}_u$, ${\bf F}_k^{(l)}$, ${\bf M}^{(l)}$, and ${\bf\Phi}^{(l)}$;
  \end{algorithmic}
\end{algorithm}

\begin{remark}
The optimization framework is general and can handle different architectures by appropriate constraints on ${\bf M}$ and choice of ${\bf V}$. In particular, letting $N_{RF}=N_B$, and  ${\bf M}={\bf V}={\bf I}_{N_{RF}}$, the problem reduces to classical digital beamforming based DFRC. Further, letting  ${\bf M}={\bf I}$ and different choice of ${\bf V}$ leads to different hybrid analog digital beamforming architectures, including partial and fully connected.   
\end{remark}

\subsection{Computational complexity}
The overall computational burden of \textbf{Algorithm~\ref{alg:AM}} is linear with the number of outer iterations. Meanwhile, at each outer iteration, the closed-form solution of radar filter ${\bf w}_{k,t}, t=1,\cdots,T$ and receive combiner ${\bf w}_{k,u}, u=1,\cdots,U$ is given by solving generalized Rayleigh quotient problem with the complexity of $\mathcal{O}(T{N_B}^2+U{N_U}^2)$. 
Then, for the update of the transmit beamforming matrix ${\bf F}_k, k=1,\cdots,K$, the computational cost of \textbf{Algorithm \ref{alg:DBS}} is linear with the number of inner iterations $l_2$. At each inner iteration of the Dinkelbach-based method, the problem is solved by the CVX \cite{grant2009cvx} with the complexity of $\mathcal{O}(K^3{N^{3}_{RF}}U^{3})$. Similarly, for the update of the holographic beamforming matrix ${\bf M}$, the complexity is $\mathcal{O}({N^{3}_{B}})$
In order to update the phase-shift matrix ${\bf\Phi}$ in \textbf{Algorithm \ref{alg:C-ADMM}}, the C-ADMM and RSD algorithm are combined with the total complexity $\mathcal{O}(l_4(2l_3{N^{2}_R}+{N_R^2)})$, where $l_3$ and $l_4$ denote the maximum iteration number of RSD and C-ADMM, respectively. Finally, the total complexity of the proposed algorithm is $\mathcal{O}({T N^2_{B}}+{U N^2_{U}}+K^3{N^{3}_{RF}}U^{3}+{N^{3}_{B}}+l_4(2l_3{N^2_R}+{N^2_R}))$ for each outer iteration. 



\section{Numerical Experiments}
\label{sec:Sims}
We validated our models and methods through extensive numerical experiments. 
Unless otherwise specified, in the simulations, RHS, RIS, and MU are equipped with the square UPA with $N_B=25$, $N_R=100$, and $N_U=16$ elements, respectively. Meanwhile, DFBS equips $N_{RF}=4$ feeds, which are connected with RHS. We deploy a DFBS at the 3-D position ${\bf p}_B\!=\![0\,\mathrm{m},0\,\mathrm{m},0\,\mathrm{m}]$ and a RIS at ${\bf p}_R\!=\![5\,\mathrm{m},5\,\mathrm{m},5\,\mathrm{m}]$. Two targets are located at ${\bf p}_T(1)\!=\![1\,\mathrm{m},2\,\mathrm{m},3\,\mathrm{m}]$, and ${\bf p}_T(2)\!=\![2\,\mathrm{m},1\,\mathrm{m},1\,\mathrm{m}]$, three clutter discretes are at ${\bf p}_C(1)\!=\![2.4\,\mathrm{m},3.4\,\mathrm{m},3.8\,\mathrm{m}]$, ${\bf p}_C(2)\!=\![3.2\,\mathrm{m},2.8\,\mathrm{m},2.8\,\mathrm{m}]$, and ${\bf p}_C(3)\!=\![5.6\,\mathrm{m}, 3.8\,\mathrm{m}, 2.0\,\mathrm{m}]$, respectively. Meanwhile, we consider three users ($U=3$) located at ${\bf p}_U(1)\!=\![-6\,\mathrm{m},1.5\,\mathrm{m},3\,\mathrm{m}]$, ${\bf p}_U(2)\!=\![-5\,\mathrm{m},1.5\,\mathrm{m},3\,\mathrm{m}]$, and ${\bf p}_U(3)\!=\![1\,\mathrm{m},2\,\mathrm{m},2.5\,\mathrm{m}]$, respectively. The central frequency of the wideband DFRC is $f_c=0.15$ THz and the subcarrier spacing  of OFDM is set to $\triangle{f}=0.5$ GHz. In the existing and emerging communications standards (LTE, LTE/A and 5G NR), a physical resource block comprising 12 subcarriers and a number of OFDM symbols forms the basis for resource allocation. Motivated by this, we consider the subcarrier number $K=16$ in our study. The Rician factor and number of NLoS path for the RHS-user link and RIS-user link are set as ${\Upsilon}_{C_u}^{\mathrm{dir}}={\Upsilon}_{C_u}^{\mathrm{RIS}}=100$ and $L_d=L_r=15$; See \cite{cheng2021hybrid}.
The inter-element spacing for the RHS, RIS, and MU are set as $\lambda_c/6$, $\lambda_c/2$ and $\lambda_c/2$, respectively, alone with both $x$- and $y$-axis. Finally, the refractive index of the RHS is set to $\gamma=\sqrt{3}$; See \cite{zhang2022holographic}.

As indicated earlier, we consider the distance-dependent path loss model $g = \sqrt{K_0(\frac{r_0}{r})^{\epsilon}}$ for both radar and communications paths. 
The signal attenuation is set as $K_0=-30\mathrm{dB}$ at the reference distance $r_0=1\,\mathrm{m}$. The relative distance of DFBS-$t$-th target, RIS-$t$-th target, DFBS-$q$-th clutter, RIS-$q$-th clutter, DFBS-$u$-th user, RIS-$u$-th user and  DFBS-RIS,  are, respectively, given by 
\begin{subequations}\label{distance}
  \begin{align}
     & r_{Bt} = \|{\bf p}_B-{\bf p}_T(t)\|_2, ~~~~ r_{Rt} = \|{\bf p}_R-{\bf p}_T(t)\|_2, \\
     & r_{Bq} = \|{\bf p}_B-{\bf p}_C(q)\|_2,  ~~~ r_{Rq} = \|{\bf p}_R-{\bf p}_C(q)\|_2, \\  
     & r_{Bu}  = \|{\bf p}_B-{\bf p}_U(u)\|_2,  ~~~ r_{Ru} = \|{\bf p}_R-{\bf p}_U(u)\|_2, \\ 
     & r_{BR} = \|{\bf p}_B-{\bf p}_R\|_2,  \quad  ~~~~  \forall t, ~ \forall q, \mathrm{and} ~ \forall u.    
  \end{align}
\end{subequations}
The corresponding path loss exponents are given by $\epsilon_{BR}=2.0$, $\epsilon_{\mathrm{dir}}=\epsilon_{Bt}=\epsilon_{Bq}=2.4$, $\epsilon_{Bu}=2.8$, $\epsilon_{Rt}= \epsilon_{Rq}=\epsilon_{Ru}=2.0, ~\forall t,\forall q$ and $\forall u$ as a representative of weak and strong line-of-sight links. Based on above, the fading component, i.e., channel gain for DFBS-target $(g_{Bt, k})$, RIS-target $(g_{Rt, k})$, DFBS-clutter $(g_{Bq, k})$, RIS-clutter $(g_{Rq, k})$, DFBS-user $(g_{Bu, k})$, RIS-user $(g_{Ru, k})$, and DFBS-RIS $(g_{BR, k})$ are obtained by appropriate substitution for $\epsilon$ and $r$ in the expression $\sqrt{K_0(\frac{r_0}{r})^{\epsilon}}$.
%
Without loss the generality, we set the RCS of targets and clutters as ${\alpha_t}=1$ and ${\alpha_q}=1$, respectively. The transmit power at each subcarrier is set to $\mathcal{P}_k\!=\!5\,\mathrm{dBw},\forall k$ and the noise variances are set to $\sigma_R^2\!=\!-45\,\mathrm{dBm}$ and $\sigma_C^2\!=\!-55\,\mathrm{dBm}$ for radar and communication, respectively. We set the SINR threshold  $\eta=9\,\mathrm{dB}$ for all users. The step size for RSD algorithm is set to $3.98$ and the termination threshold for AM algorithm is set as $\zeta_3=10^{-4}$. The initialization of ${\bf\Phi}^{(0)}$ is randomly generated diagonal matrix, whose entries are assumed be constant modulus with random phase-shifts. The maximum iteration for RSD, C-ADMM and AM are set as $100$, $24$ and $30$. 
\vspace{-0.2cm}
\subsection{Convergence of the Proposed Algorithm }


\begin{figure*}[!t]\centering
\vspace{-0.5cm}  \subfloat[]{
    \includegraphics[width=0.5\textwidth]{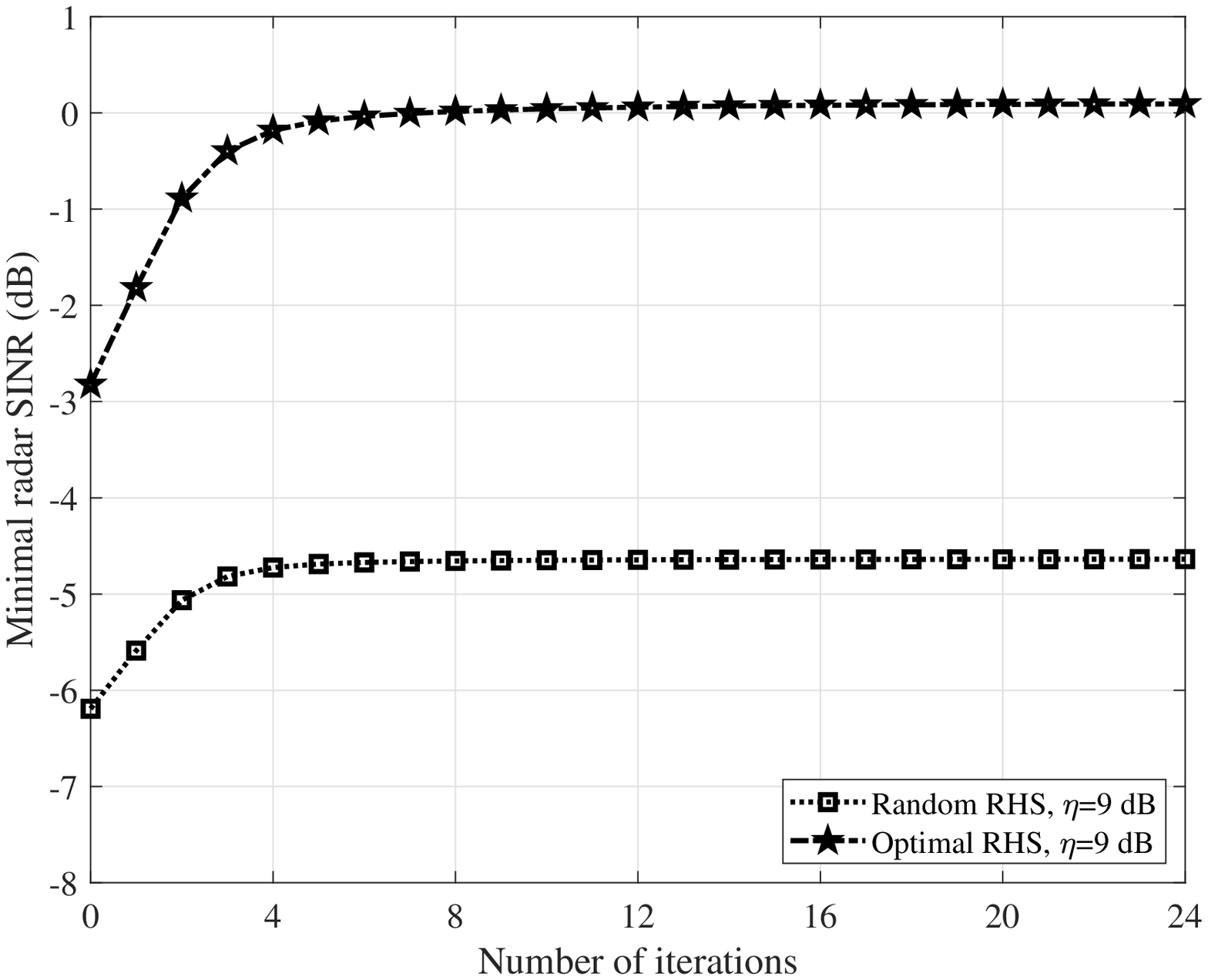}
    }
  \subfloat[]{
    \includegraphics[width=0.5\textwidth]{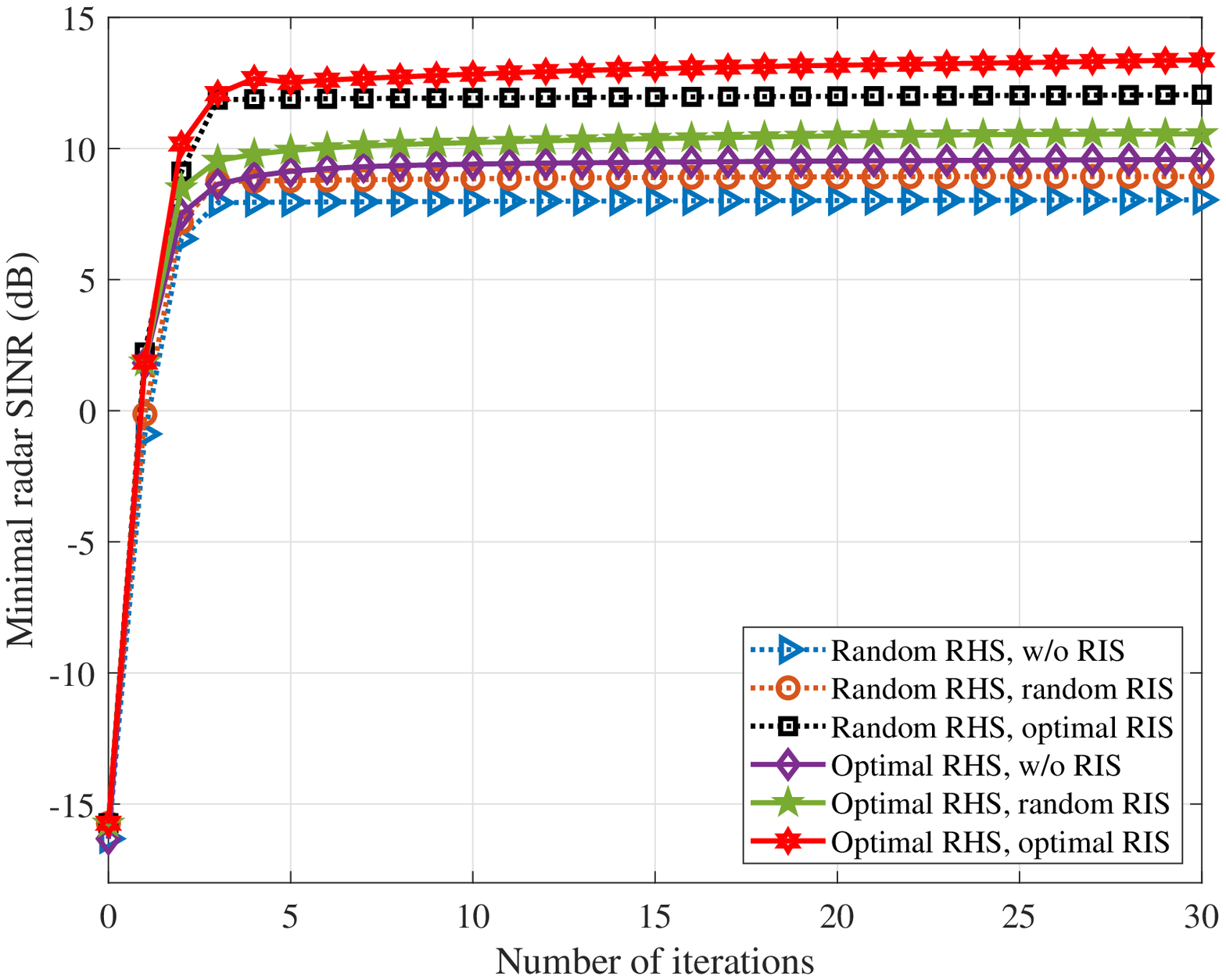}
    }  
\vspace{-0.2cm}\caption{Convergence performance of the proposed algorithm;  (a) Minimum radar SINR versus number of iterations for C-ADMM;   (b) Minimum radar SINR versus number of iterations for AM.}
\vspace{-0.8cm}\label{fig:convergence}
\end{figure*}

Fig.~\ref{fig:convergence}(a) illustrates the convergence of C-ADMM algorithm for solving the subproblem related to transmit beamformer design in the first AM iteration. We compare two different scenarios: (1). {\em Random RHS}, which keep ${\bf M}$ fixed before C-ADMM; (2). {\em Optimal RHS}, which update ${\bf M}$ by \textbf{Algorithm \ref{alg:DBS}} before C-ADMM. Note that even though the RSD algorithm can not obtain a close-form solution for the nonconvex manifold optimization problem in each ADMM iteration, the objective value of the proposed C-ADMM is still monotonically increasing in the first AM iteration which indicates the phase-shift update provides additional enhancement of minimum radar SINR. Meanwhile, the optimal RHS case reaps the better radar SINR performance.    


Fig.~\ref{fig:convergence}(b) demonstrates the overall convergence  of proposed AM algorithm in different scenarios: (1). {\em Random RHS, w/o RIS} involves optimal receive filter, digital beamforming, random RHS, and non-RIS, which is set as the {\em benchmark}; (2). {\em Random RHS, random RIS} with optimal receive filter, digital beamforming, random RHS, and random RIS; (3). {\em Random RHS, optimal RIS} comprises optimal receive filter, digital beamforming, random RHS, and optimal RIS; (4). {\em Optimal RHS, w/o RIS} involves optimal receive filter, digital beamforming, RHS, and non-RIS; (5). {\em Optimal RHS, random RIS} with optimal receive filter, digital beamforming, RHS, and random RIS; (6). {\em Optimal RHS, optimal RIS} comprises optimal receive filter, digital beamforming, RHS, and RIS.
Despite multiple linearization steps being utilized for both objective  and constraint functions in the subproblems, the proposed AM algorithm is convergent in about $10$ iterations. The  minimum radar SINR objective is monotonically increasing with the iterations since the objective value of the original problem is always greater than the linearized problem and Dinkelbach method can also guarantee the monotonic objective value. The cases with the optimal RHS achieve the higher worst-case radar SINR compared with the random RHS. Meanwhile, with random phase-shifts for RIS, the worst-case radar SINR is lower than  that the  optimal phase-shifts but higher than the non-RIS case. Optimizing the phase-shifts leads to at least 2.8 and 3.7 dB radar SINR gain for random RIS and non-RIS deployment, respectively.  

\vspace{-0.2cm}
\subsection{Effect of Transmit Power} 
 
\begin{figure}[t]
\centering{\includegraphics[width=0.5\columnwidth]{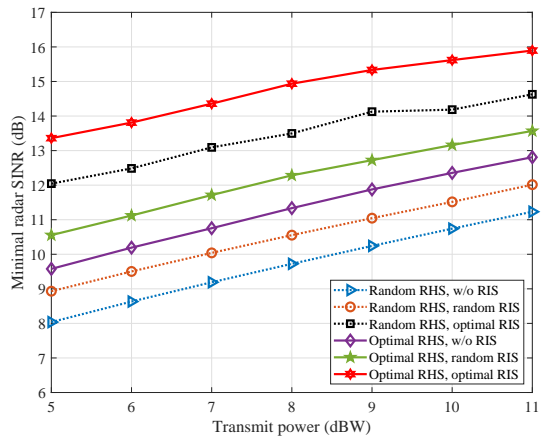}}
\caption{Minimum Radar SINR versus transmit power, $\epsilon_{\mathrm{dir}}=2.4$, $\eta=9$ dB, $N_B=16$, $N_R=100$.
\label{Fig3R}}
\end{figure}

We illustrate the variation in the minimum achievable radar SINR with respect to different parameters to demonstrate the flexibility of our approach. Fig.~\ref{Fig3R} shows the achievable worst-case radar SINR versus the $\mathcal{P}_k$ for each subcarrier.  Clearly, if we simultaneously optimize the receive filter, digital beamformer, holographic beamformer and passive beamformer, the highest radar SINR is achieved compared with random RHS, non-RIS and random RIS.
Furthermore, for the RIS-assisted DFRC, the optimization of the holographic beamforming by RHS can provide around 1.3 dB SINR gain owing to its flexibility on radiation amplitude controlling. 
It is also observed that for all six cases (i.e., random RHS, optimized RHS,  w/o RIS, random RIS and optimized RIS), increasing of 1 dB transmit power can bring around 0.4 dB radar SINR enhancement. Meanwhile, the proposed algorithm with both optimal RHS and RIS achieves the highest radar SINR which indicates the advantage of our proposed RIS-assisted holographic DFRC system.

\vspace{-0.2cm}
\subsection{Effect of radar LoS Pathloss Exponent}

\begin{figure}[t]
\centering{\includegraphics[width=0.5\columnwidth]{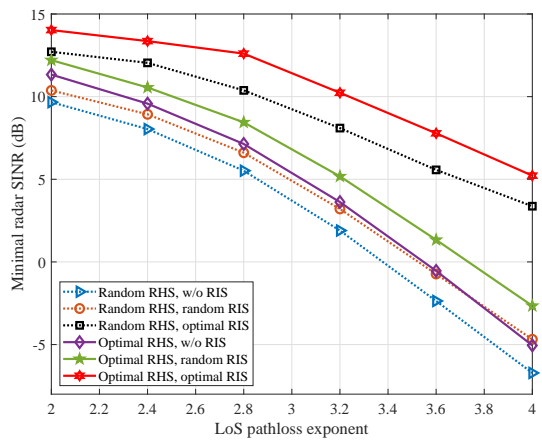}}
\caption{Minimum Radar SINR versus radar LoS pathloss exponent, $\mathcal{P}_k=5$ dBw, $\eta=9$ dB, $N_B=16$, $N_R=100$.
\label{Fig4R}}
\end{figure}

Fig.~\ref{Fig4R} shows the achievable worst-case radar SINR versus radar LoS pathloss exponent $\epsilon_{\mathrm{dir}}$ (i.e., direct pathloss exponent between RHS and targets/clutters). The radar SINR is gradually reduced with the increase in the LoS pathloss exponent. 
Meanwhile, the proposed methods with optimal RIS (random or optimal RHS) provide better radar SINR with the weak direct path, i.e., $\epsilon_{\mathrm{dir}}\geq 3.6$. 
This demonstrates RIS is more effective when a stable LoS path is missing. 

\vspace{-0.2cm}
\subsection{Effect of Surface Element Number}

\begin{figure*}[!t]\centering
\vspace{-0.5cm}  \subfloat[]{
    \includegraphics[width=0.5\textwidth]{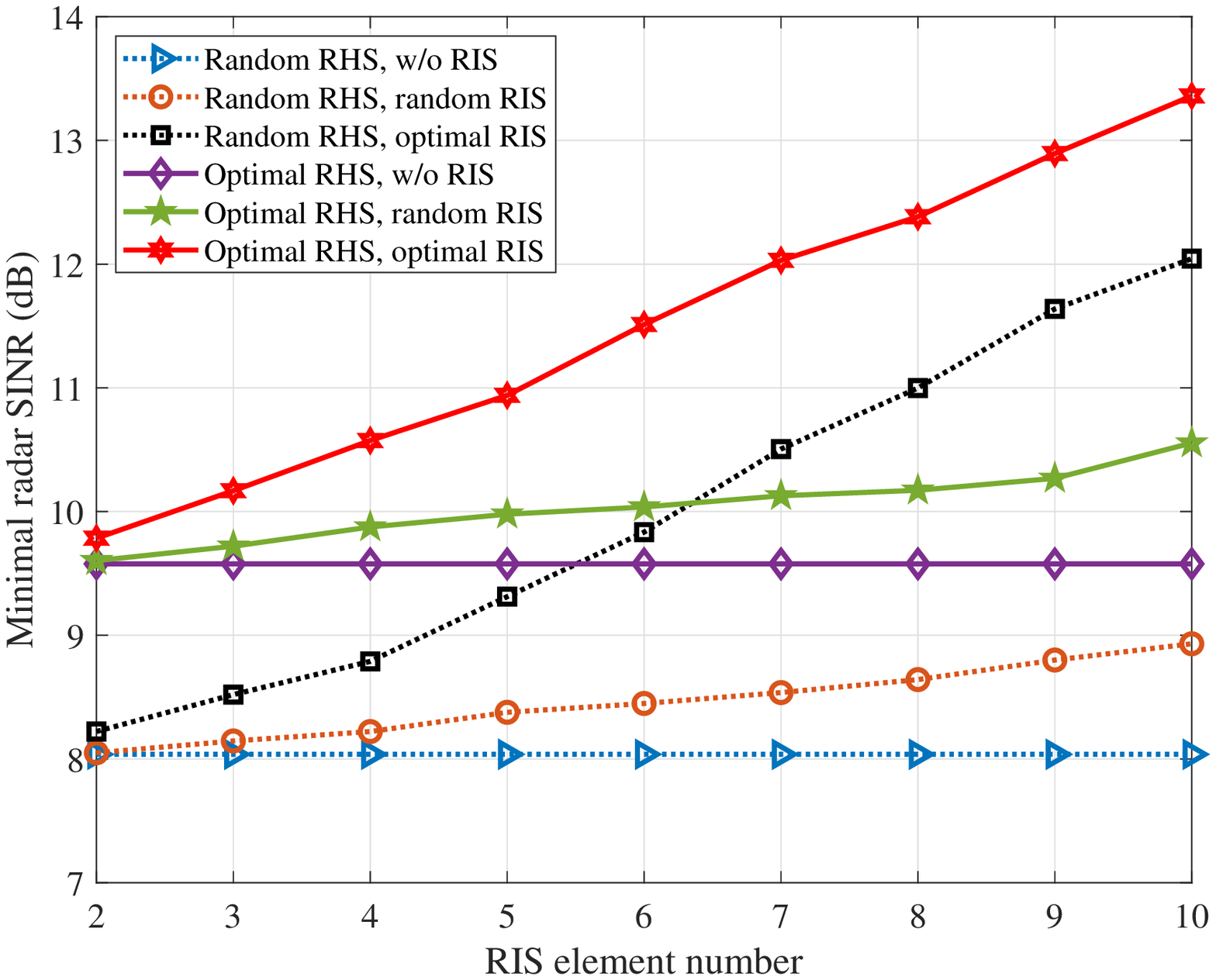}
    }
  \subfloat[]{
    \includegraphics[width=0.5\textwidth]{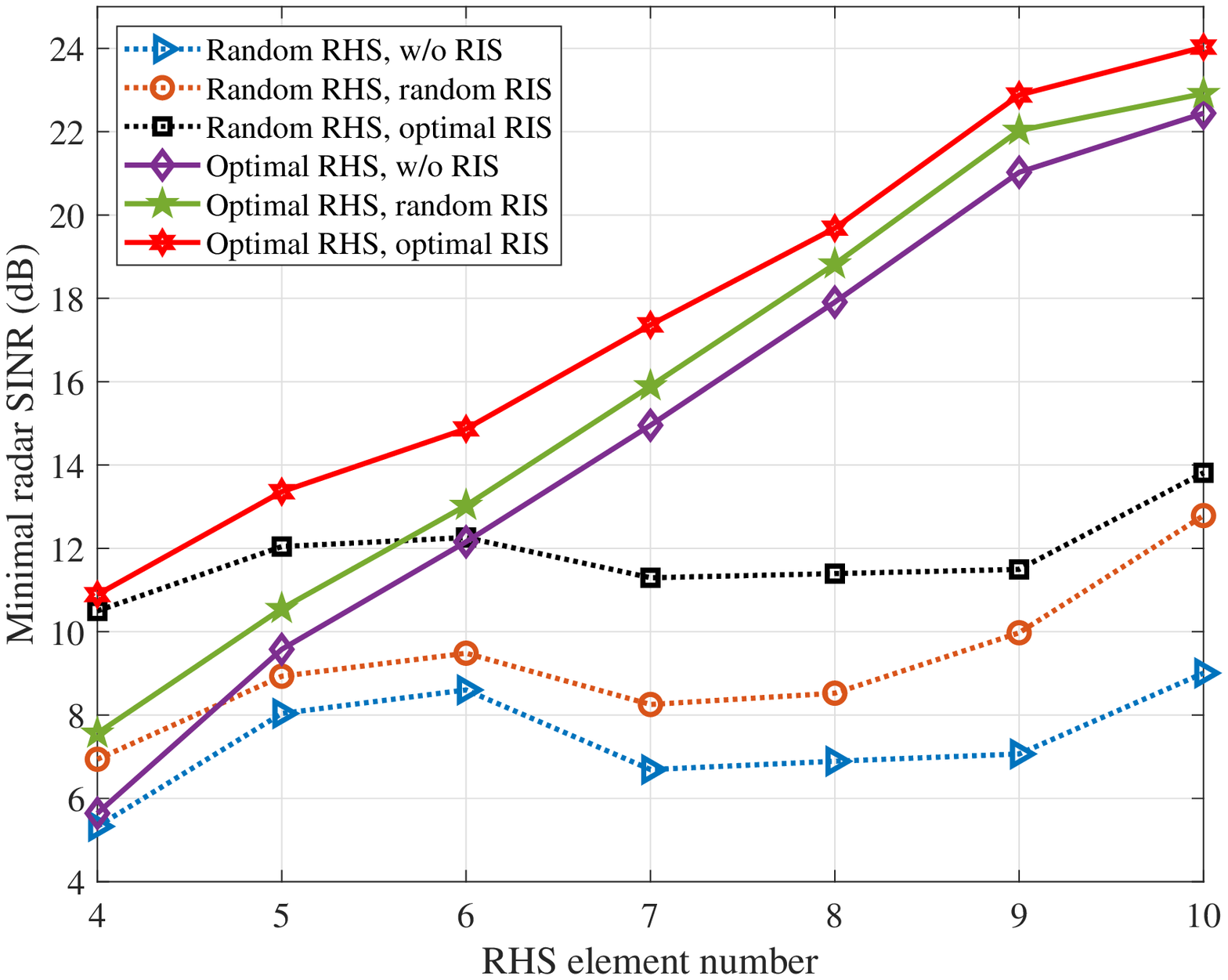}
    }  
\vspace{-0.2cm}\caption{Radar performance in terms of the element number of surfaces;  (a) Minimum Radar SINR versus RIS element number $N_x^R=N_y^R$, $\mathcal{P}_k=5$ dBw, $\eta=9$ dB, $\epsilon_{\mathrm{dir}}=2.4$, $N_B=16$;   (b) Minimum Radar SINR versus  RHS element number $N_x^B=N_y^B$, $\mathcal{P}_k=5$ dBw, $\eta=9$ dB, $\epsilon_{\mathrm{dir}}=2.4$, $N_R=100$.}
\vspace{-0.8cm}\label{fig:surface}
\end{figure*}



Fig.~\ref{fig:surface}(a), shows the achievable worst-case radar SINR versus RIS element number w.r.t $x$- (or $y$-) axis, $N_x^R$ (or $N_y^R$). It is observed that random RIS can not provide a stable performance enhancement compared with the optimal RIS, even with the larger element number. Different from that, the proposed method with the optimal RHS and RIS is able to enhance the system continuously. Meanwhile, it is seen that each passive RIS element enhancement w.r.t $x$- (or $y$-) axis can offer around 0.4 dB SINR improvement under the optimal receive filter, digital beamforming, holographic beamforming, and passive beamforming, see \textit{red curve}. This kind of enhancement is crucial for DFRC system, especially in the dense environment, where the LoS components are weak. Fig.~\ref{fig:surface}(b) shows the achievable worst-case radar SINR versus RHS element number w.r.t $x$- (or $y$-) axis, $N_x^B$ (or $N_y^B$). 
With the increase in RHS elements, random RHS does not provide a stable radar SINR improvement. On the contrary, the optimal RHS is able to enhance the system continuously which highlights the importance of proposed joint design scheme for RIS-assisted holographic DFRC.
  
\vspace{-0.2cm}
\subsection{Effect of Communication SINR}

\begin{figure}[t]
\centering{\includegraphics[width=0.5\columnwidth]{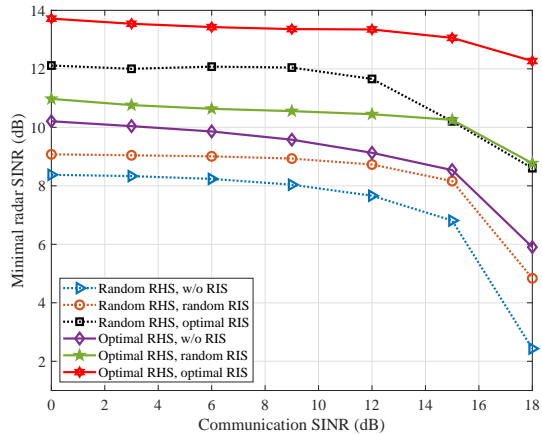}}
\caption{Minimum Radar SINR versus communication SINR threshold, $\mathcal{P}_k=5$ dBw, $\epsilon_{\mathrm{dir}}=2.4$, $N_B=16$, $N_R=100$.
\label{Fig7R}}
\end{figure}

In Fig.~\ref{Fig7R}, the results of the achievable worst-case radar SINR versus $\eta$ are presented. We vary the  communications SINR threshold from 0 to 18 dB which covers the demands in most existing wireless applications. We observe that, in the DFRC system, the higher requirement on the communications SINR leads to the performance loss for radar SINR. Hence, even with the assistance of RIS, the proposed holographic DFRC system still demonstrates a performance trade-off between radar and communications, albeit, an enhanced one. Note that the threshold of communications SINR can be flexibly selected based on the system requirement, e.g., decoding performance or outage probability, and it can be also varied for different users. 

\subsection{Effect of Number of Users} 
\begin{figure}[t]
\centering{\includegraphics[width=0.5\columnwidth]{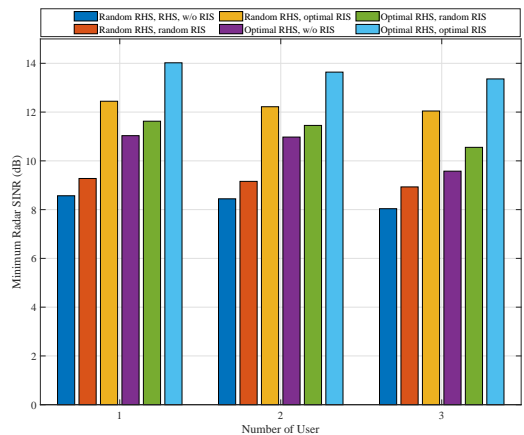}}
\caption{Minimum Radar SINR versus number of user, $\mathcal{P}_k=5$ dBw, $\epsilon_{\mathrm{dir}}=2.4$, $\eta=9$ dB, $N_B=16$, $N_R=100$.
\label{Fig8R}}
\end{figure}
Fig.~\ref{Fig8R} depicts the achievable worst-case radar SINR as a function of $U$. It  again indicates that the proposed method with the optimal RIS and RHS achieves the best radar SINR compared with non-RIS, random-RIS and random-RHS cases in terms of different user numbers. On the other side, the increase of one user can lead to only around 0.35 dB radar SINR loss which expresses the robustness of the proposed DFRC system.

\section{Summary}
\label{sec:summ}
We considered the joint deployment of the RHS and RIS to assist a wideband DFRC system with OFDM signaling. Our design of digital, holographic, and passive beamformers shows improvement in the performance of the DFRC system
when compared with non-RIS and non-RHS systems. The key challenge to the design problem arises from the coupling of various parameters and nonconvexity. We showed that our alternating optimization approach facilitates not only decoupling but also a tractable design.

For highly dynamic wideband channels, machine learning methods may be employed to estimate the channel state \cite{elbir2021terahertz}. This may also be incorporated with the RHS DFRC systems. A particularly complicated procedure in the RHS is estimation of angle-of-arrivals because, unlike phased arrays whose feeds directly receive the signals, RHS feeds receive the signals after modulation by holographic patterns. This requires an additional maximum likelihood estimation step for AoA estimation.

Holographic DFRC is currently at an early stage of research. As a result, substantial challenges in prototyping, channel modeling, and optimized design remain. Graphene-based arrays and leaky-wave antennas are other alternatives for realizing these antenna structures \cite{elbir2022terahertz}. Further, exploiting other non-OFDM multiple access technologies for signaling also offers a promising research avenue for holographic DFRC in the near future \cite{zhang2022holographic}.

\appendices
\section{Proof of the Proposition~\ref{cor:opt}}
The maximization problem of the generalized Rayleigh quotient 
\begin{align}\label{GRQ_p1}
    \mathop{{\mathrm{maximize}}}\limits_{{\bf w}\neq{\bf 0}} & \quad \frac{{\bf w}^H{\bf A}{\bf w}}{{\bf w}^H{\bf B}{\bf w}} 
\end{align}
can be equivalently reformulated as 
  \begin{align}  \label{homo_QCQP}
    \mathop{{\mathrm{maximize}}}\limits_{{\bf w}\neq{\bf 0}} \quad {\bf w}^H{\bf A}{\bf w},{\quad} 
         \textrm{subject to}  \quad {\bf w}^H{\bf B}{\bf w}=1.
   \end{align}
Note that problem \eqref{homo_QCQP} is the complex-valued homogeneous QCQP, which can not be directly solved due to the convex objective function and quadratic constraint. However, we can write the Lagrangian function of problem \eqref{homo_QCQP} as 
\begin{align}\label{Lagrangian_QCQP}
    \mathcal{L}({\bf w},\lambda)={\bf w}^H{\bf A}{\bf w}-\lambda({\bf w}^H{\bf B}{\bf w}-1),
\end{align}
where $\lambda$ denotes the corresponding Lagrange multiplier. Setting the derivative of the Lagrangian in \eqref{Lagrangian_QCQP} with respect to ${\bf w}$ to zero, i.e.,
\begin{align}\label{derivarion_QCQP}
    \frac{\partial\mathcal{L}({\bf w},\lambda)}{\partial{\bf w}}=2{\bf A}{\bf w}-2\lambda{\bf B}{\bf w}={\bf 0}.
\end{align}
Based on above, the extreme value of equation \eqref{Lagrangian_QCQP} and should satisfy  
\begin{align}\label{equ_QCQP}
    {\bf B}^{-1}{\bf A}{\bf w}=\lambda{\bf w}.
\end{align}
According to \eqref{equ_QCQP}, we found that $\lambda$ and ${\bf w}$ are the eigenvalue and corresponding eigenvector of ${\bf B}^{-1}{\bf A}$, respectively. Meanwhile, from \eqref{homo_QCQP} and \eqref{derivarion_QCQP}, we have 
\begin{align}\label{equality}
    {\bf w}^H{\bf A}{\bf w}=\lambda{\bf w}^H{\bf B}{\bf w}=\lambda.
\end{align} 
Hence, we conclude that the maximum values $\lambda^{\ast}$ of ${\bf w}^H{\bf A}{\bf w}$ and the generalized Rayleigh quotient \eqref{GRQ_p1} is $\lambda^{\ast}=\lambda_{\mathrm{max}}({\bf B}^{-1}{\bf A})$, and the corresponding vector is ${\bf w}^{\ast}=\rho_{\mathrm{max}}({\bf B}^{-1}{\bf A})$. This completes the proof.

\ifCLASSOPTIONcaptionsoff
  \newpage
\fi

\bibliographystyle{IEEEtran}
\bibliography{ref}

\begin{thebibliography}{10}
\providecommand{\url}[1]{#1}
\csname url@samestyle\endcsname
\providecommand{\newblock}{\relax}
\providecommand{\bibinfo}[2]{#2}
\providecommand{\BIBentrySTDinterwordspacing}{\spaceskip=0pt\relax}
\providecommand{\BIBentryALTinterwordstretchfactor}{4}
\providecommand{\BIBentryALTinterwordspacing}{\spaceskip=\fontdimen2\font plus
\BIBentryALTinterwordstretchfactor\fontdimen3\font minus
  \fontdimen4\font\relax}
\providecommand{\BIBforeignlanguage}[2]{{%
\expandafter\ifx\csname l@#1\endcsname\relax
\typeout{** WARNING: IEEEtran.bst: No hyphenation pattern has been}%
\typeout{** loaded for the language `#1'. Using the pattern for}%
\typeout{** the default language instead.}%
\else
\language=\csname l@#1\endcsname
\fi
#2}}
\providecommand{\BIBdecl}{\relax}
\BIBdecl

\bibitem{hodge2020intelligent}
J.~A. Hodge, K.~V. Mishra, and A.~I. Zaghloul, ``Intelligent time-varying
  metasurface transceiver for index modulation in 6{G} wireless networks,''
  \emph{IEEE Antennas and Wireless Propagation Letters}, vol.~19, no.~11, pp.
  1891--1895, 2020.

\bibitem{mishra2022machine}
K.~V. Mishra, A.~M. Elbir, and A.~I. Zaghloul, ``Machine learning for
  metasurfaces design and their applications,'' in \emph{Advances in
  Electromagnetics Empowered by Machine Learning}, ser. Electromagnetic Wave
  Theory and Applications, D.~H. Werner and S.~D. Campbell, Eds.\hskip 1em plus
  0.5em minus 0.4em\relax Wiley-IEEE Press, 2022, in press.

\bibitem{wu2019towards}
Q.~Wu and R.~Zhang, ``Towards smart and reconfigurable environment:
  {I}ntelligent reflecting surface aided wireless network,'' \emph{IEEE
  Communications Magazine}, vol.~58, no.~1, pp. 106--112, 2019.

\bibitem{sievenpiper1999high}
D.~Sievenpiper, L.~Zhang, R.~F. Broas, N.~G. Alexopolous, E.~Yablonovitch
  \emph{et~al.}, ``High-impedance electromagnetic surfaces with a forbidden
  frequency band,'' \emph{IEEE Transactions on Microwave Theory and
  techniques}, vol.~47, no.~11, pp. 2059--2074, 1999.

\bibitem{zhu2013linear}
H.~Zhu, S.~Cheung, K.~L. Chung, and T.~I. Yuk, ``Linear-to-circular
  polarization conversion using metasurface,'' \emph{IEEE Transactions on
  Antennas and Propagation}, vol.~61, no.~9, pp. 4615--4623, 2013.

\bibitem{minatti2015modulated}
G.~Minatti, M.~Faenzi, E.~Martini, F.~Caminita, P.~De~Vita,
  D.~Gonz{\'a}lez-Ovejero, M.~Sabbadini, and S.~Maci, ``Modulated metasurface
  antennas for space: Synthesis analysis and realizations,'' \emph{IEEE
  Transactions on Antennas and Propagation}, vol.~63, no.~4, pp. 1288--1300,
  2015.

\bibitem{mishra2018reconfigurable}
K.~V. Mishra, J.~A. Hodge, and A.~I. Zaghloul, ``Reconfigurable metasurfaces
  for radar and communications systems,'' in \emph{URSI Asia-Pacific Radio
  Science Conference}, 2019, pp. 1--4.

\bibitem{glybovski2016metasurfaces}
S.~B. Glybovski, S.~A. Tretyakov, P.~A. Belov, Y.~S. Kivshar, and C.~R.
  Simovski, ``Metasurfaces: {F}rom microwaves to visible,'' \emph{Physics
  Reports}, vol. 634, pp. 1--72, 2016.

\bibitem{esmaeilbeig2022irs}
Z.~Esmaeilbeig, K.~V. Mishra, and M.~Soltanalian, ``{IRS}-aided radar:
  {E}nhanced target parameter estimation via intelligent reflecting surfaces,''
  in \emph{IEEE Sensor Array and Multichannel Signal Processing Workshop},
  2022, pp. 286--290.

\bibitem{renzo2020smart}
M.~Di~Renzo, A.~Zappone, M.~Debbah, M.-S. Alouini, C.~Yuen, J.~de~Rosny, and
  S.~Tretyakov, ``Smart radio environments empowered by reconfigurable
  intelligent surfaces: How it works, state of research, and the road ahead,''
  \emph{IEEE Journal on Selected Areas in Communications}, vol.~38, no.~11, pp.
  2450--2525, 2020.

\bibitem{wu2019intelligent}
Q.~Wu and R.~Zhang, ``Intelligent reflecting surface enhanced wireless network
  via joint active and passive beamforming,'' \emph{IEEE Transactions on
  Wireless Communications}, vol.~18, no.~11, pp. 5394--5409, 2019.

\bibitem{buzzi2022foundations}
S.~Buzzi, E.~Grossi, M.~Lops, and L.~Venturino, ``Foundations of {MIMO} radar
  detection aided by reconfigurable intelligent surfaces,'' \emph{IEEE
  Transactions on Signal Processing}, vol.~70, pp. 1749--1763, 2022.

\bibitem{tang2021wireless}
W.~Tang, M.~Z. Chen, X.~Chen, J.~Y. Dai, Y.~Han, M.~Di~Renzo, Y.~Zeng, S.~Jin,
  Q.~Cheng, and T.~J. Cui, ``Wireless communications with reconfigurable
  intelligent surface: Path loss modeling and experimental measurement,''
  \emph{IEEE Transactions on Wireless Communications}, vol.~20, no.~1, pp.
  421--439, 2021.

\bibitem{mishra2022optm3sec}
K.~V. Mishra, A.~Chattopadhyay, S.~S. Acharjee, and A.~P. Petropulu,
  ``{OptM3Sec}: {O}ptimizing multicast {IRS}-aided multiantenna {DFRC} secrecy
  channel with multiple eavesdroppers,'' in \emph{IEEE International Conference
  on Acoustics, Speech and Signal Processing}, 2022, pp. 9037--9041.

\bibitem{watson2019non}
B.~Watson and J.~R. Guerci, \emph{Non-line-of-sight radar}.\hskip 1em plus
  0.5em minus 0.4em\relax Artech House, 2019.

\bibitem{aubry2021reconfigurable}
A.~Aubry, A.~De~Maio, and M.~Rosamilia, ``Reconfigurable intelligent surfaces
  for {N-LOS} radar surveillance,'' \emph{IEEE Transactions on Vehicular
  Technology}, vol.~70, no.~10, pp. 10\,735--10\,749, 2021.

\bibitem{abeywickrama2020intelligent}
S.~Abeywickrama, R.~Zhang, Q.~Wu, and C.~Yuen, ``Intelligent reflecting
  surface: {Practical} phase shift model and beamforming optimization,''
  \emph{IEEE Transactions on Communications}, vol.~68, no.~9, pp. 5849--5863,
  2020.

\bibitem{Ur2021joint}
H.~Ur~Rehman, F.~Bellili, A.~Mezghani, and E.~Hossain, ``Joint active and
  passive beamforming design for {IRS}-assisted multi-user {MIMO} systems: {A}
  {VAMP}-based approach,'' \emph{IEEE Transactions on Communications}, vol.~69,
  no.~10, pp. 6734--6749, 2021.

\bibitem{mishra2019toward}
K.~V. Mishra, M.~B. Shankar, V.~Koivunen, B.~Ottersten, and S.~A. Vorobyov,
  ``Toward millimeter-wave joint radar communications: {A} signal processing
  perspective,'' \emph{IEEE Signal Processing Magazine}, vol.~36, no.~5, pp.
  100--114, 2019.

\bibitem{wei2022quantized}
T.~Wei, L.~Wu, K.~V. Mishra, and M.~R.~B. Shankar, ``Irs-aided wideband
  dual-function radar-communications with quantized phase-shifts,'' in
  \emph{2022 IEEE 12th Sensor Array and Multichannel Signal Processing Workshop
  (SAM)}, 2022, pp. 465--469.

\bibitem{elbir2022rise}
A.~M. Elbir, K.~V. Mishra, M.~R.~B. Shankar, and S.~Chatzinotas, ``The rise of
  intelligent reflecting surfaces in integrated sensing and communications
  paradigms,'' \emph{IEEE Network}, pp. 1--8, 2022.

\bibitem{wei2022multiple}
T.~Wei, L.~Wu, K.~V. Mishra, and M.~R.~B. Shankar, ``Multi-{IRS}-aided
  {D}oppler-tolerant wideband {DFRC} system,'' \emph{arXiv preprint
  arXiv:2207.02157}, 2022.

\bibitem{jiang2022intelligent}
Z.-M. Jiang, M.~Rihan, P.~Zhang, L.~Huang, Q.~Deng, J.~Zhang, and E.~M.
  Mohamed, ``Intelligent reflecting surface aided dual-function radar and
  communication system,'' \emph{IEEE Systems Journal}, vol.~16, no.~1, pp.
  475--486, 2022.

\bibitem{wang2022jointwaveform}
X.~Wang, Z.~Fei, J.~Huang, and H.~Yu, ``Joint waveform and discrete phase shift
  design for {RIS}-assisted integrated sensing and communication system under
  cram\'{e}r-{R}ao bound constraint,'' \emph{IEEE Transactions on Vehicular
  Technology}, vol.~71, no.~1, pp. 1004--1009, 2022.

\bibitem{wei2022simultaneous}
T.~Wei, L.~Wu, K.~V. Mishra, and S.~M.~R. Bhavani, ``Simultaneous
  active-passive beamformer design in {IRS}-enabled multi-carrier {DFRC}
  system,'' in \emph{European Signal Processing Conference}, 2022, pp.
  1007--1011.

\bibitem{hwang2020binary}
R.-B.~R. Hwang, ``Binary meta-hologram for a reconfigurable holographic
  metamaterial antenna,'' \emph{Scientific Reports}, vol.~10, no.~1, pp. 1--10,
  2020.

\bibitem{sleasman2015waveguide}
T.~Sleasman, M.~F. Imani, W.~Xu, J.~Hunt, T.~Driscoll, M.~S. Reynolds, and
  D.~R. Smith, ``Waveguide-fed tunable metamaterial element for dynamic
  apertures,'' \emph{IEEE Antennas and Wireless Propagation Letters}, vol.~15,
  pp. 606--609, 2015.

\bibitem{deng2021survey}
R.~Deng, B.~Di, H.~Zhang, D.~Niyato, Z.~Han, H.~V. Poor, and L.~Song,
  ``Reconfigurable holographic surfaces for future wireless communications,''
  \emph{IEEE Wireless Communications}, vol.~28, no.~6, pp. 126--131, 2021.

\bibitem{deng2022reconfigurable}
R.~Deng, B.~Di, H.~Zhang, Y.~Tan, and L.~Song, ``Reconfigurable holographic
  surface-enabled multi-user wireless communications: Amplitude-controlled
  holographic beamforming,'' \emph{IEEE Transactions on Wireless
  Communications}, vol.~21, no.~8, pp. 6003--6017, 2022.

\bibitem{gholam2011beamforming}
F.~Gholam, J.~Via, and I.~Santamaria, ``Beamforming design for simplified
  analog antenna combining architectures,'' \emph{IEEE Transactions on
  Vehicular Technology}, vol.~60, no.~5, pp. 2373--2378, 2011.

\bibitem{johnson2015sidelobe}
M.~C. Johnson, S.~L. Brunton, N.~B. Kundtz, and J.~N. Kutz, ``Sidelobe
  canceling for reconfigurable holographic metamaterial antenna,'' \emph{IEEE
  Transactions on Antennas and Propagation}, vol.~63, no.~4, pp. 1881--1886,
  2015.

\bibitem{di2021reconfigurable}
B.~Di, ``Reconfigurable holographic metasurface aided wideband {OFDM}
  communications against beam squint,'' \emph{IEEE Transactions on Vehicular
  Technology}, vol.~70, no.~5, pp. 5099--5103, 2021.

\bibitem{zeng2022reconfigurable}
S.~Zeng, H.~Zhang, B.~Di, H.~Qin, X.~Su, and L.~Song, ``Reconfigurable
  refractive surfaces: An energy-efficient way to holographic {MIMO},''
  \emph{IEEE Communications Letters}, pp. 1--1, 2022.

\bibitem{zhang2022holographic}
H.~Zhang, H.~Zhang, B.~Di, M.~D. Renzo, Z.~Han, H.~V. Poor, and L.~Song,
  ``Holographic integrated sensing and communication,'' \emph{IEEE Journal on
  Selected Areas in Communications}, vol.~40, no.~7, pp. 2114--2130, 2022.

\bibitem{dardari2021holographic}
D.~Dardari and N.~Decarli, ``Holographic communication using intelligent
  surfaces,'' \emph{IEEE Communications Magazine}, vol.~59, no.~6, pp. 35--41,
  2021.

\bibitem{wan2021terahertz}
Z.~Wan, Z.~Gao, F.~Gao, M.~D. Renzo, and M.-S. Alouini, ``Terahertz massive
  {MIMO} with holographic reconfigurable intelligent surfaces,'' \emph{IEEE
  Transactions on Communications}, vol.~69, no.~7, pp. 4732--4750, 2021.

\bibitem{elbir2021terahertz}
A.~M. Elbir, K.~V. Mishra, and S.~Chatzinotas, ``Terahertz-band joint
  ultra-massive {MIMO} radar-communications: {M}odel-based and model-free
  hybrid beamforming,'' \emph{IEEE Journal of Special Topics in Signal
  Processing}, vol.~15, no.~6, pp. 1468--1483, 2021.

\bibitem{li2021lntelligent}
H.~Li, W.~Cai, Y.~Liu, M.~Li, Q.~Liu, and Q.~Wu, ``Intelligent reflecting
  surface enhanced wideband {MIMO-OFDM} communications: {F}rom practical model
  to reflection optimization,'' \emph{IEEE Transactions on Communications},
  vol.~69, no.~7, pp. 4807--4820, 2021.

\bibitem{yang2020dual}
J.~Yang, G.~Cui, X.~Yu, and L.~Kong, ``Dual-use signal design for radar and
  communication via ambiguity function sidelobe control,'' \emph{IEEE
  Transactions on Vehicular Technology}, vol.~69, no.~9, pp. 9781--9794, 2020.

\bibitem{alhujaili2019transmit}
K.~Alhujaili, V.~Monga, and M.~Rangaswamy, ``Transmit {MIMO} radar beampattern
  design via optimization on the complex circle manifold,'' \emph{IEEE
  Transactions on Signal Processing}, vol.~67, no.~13, pp. 3561--3575, 2019.

\bibitem{nai2010beampattern}
S.~E. Nai, W.~Ser, Z.~L. Yu, and H.~Chen, ``Beampattern synthesis for linear
  and planar arrays with antenna selection by convex optimization,'' \emph{IEEE
  Transactions on Antennas and Propagation}, vol.~58, no.~12, pp. 3923--3930,
  2010.

\bibitem{wang2019overview}
M.~Wang, F.~Gao, S.~Jin, and H.~Lin, ``An overview of enhanced massive {MIMO}
  with array signal processing techniques,'' \emph{IEEE Journal of Selected
  Topics in Signal Processing}, vol.~13, no.~5, pp. 886--901, 2019.

\bibitem{vlachos2019wideband}
E.~Vlachos, G.~C. Alexandropoulos, and J.~Thompson, ``Wideband {MIMO} channel
  estimation for hybrid beamforming millimeter wave systems via random spatial
  sampling,'' \emph{IEEE Journal of Selected Topics in Signal Processing},
  vol.~13, no.~5, pp. 1136--1150, 2019.

\bibitem{shtaiwi2021channel}
E.~Shtaiwi, H.~Zhang, S.~Vishwanath, M.~Youssef, A.~Abdelhadi, and Z.~Han,
  ``Channel estimation approach for {RIS} assisted {MIMO} systems,'' \emph{IEEE
  Transactions on Cognitive Communications and Networking}, vol.~7, no.~2, pp.
  452--465, 2021.

\bibitem{yang2020intelligent}
Y.~Yang, B.~Zheng, S.~Zhang, and R.~Zhang, ``Intelligent reflecting surface
  meets ofdm: Protocol design and rate maximization,'' \emph{IEEE Transactions
  on Communications}, vol.~68, no.~7, pp. 4522--4535, 2020.

\bibitem{zhang2021joint}
Z.~Zhang and L.~Dai, ``A joint precoding framework for wideband reconfigurable
  intelligent surface-aided cell-free network,'' \emph{IEEE Transactions on
  Signal Processing}, vol.~69, pp. 4085--4101, 2021.

\bibitem{zhang2020capacity}
S.~Zhang and R.~Zhang, ``Capacity characterization for intelligent reflecting
  surface aided mimo communication,'' \emph{IEEE Journal on Selected Areas in
  Communications}, vol.~38, no.~8, pp. 1823--1838, 2020.

\bibitem{wu2022wideband}
L.~Wu, K.~Lou, J.~Ke, J.~Liang, Z.~Luo, J.~Y. Dai, Q.~Cheng, and T.~J. Cui, ``A
  wideband amplifying reconfigurable intelligent surface,'' \emph{IEEE
  Transactions on Antennas and Propagation}, vol.~70, no.~11, pp.
  10\,623--10\,631, 2022.

\bibitem{xu2023bandwidth}
Z.~Xu and A.~Petropulu, ``A bandwidth efficient dual-function radar
  communication system based on a {MIMO} radar using {OFDM} waveforms,''
  \emph{IEEE Transactions on Signal Processing}, vol.~71, pp. 401--416, 2023.

\bibitem{wu2017transmit}
L.~Wu, P.~Babu, and D.~P. Palomar, ``Transmit waveform/receive filter design
  for mimo radar with multiple waveform constraints,'' \emph{IEEE Transactions
  on Signal Processing}, vol.~66, no.~6, pp. 1526--1540, 2017.

\bibitem{tsinos2021joint}
C.~G. Tsinos, A.~Arora, S.~Chatzinotas, and B.~Ottersten, ``Joint transmit
  waveform and receive filter design for dual-function radar-communication
  systems,'' \emph{IEEE Journal of Selected Topics in Signal Processing},
  vol.~15, no.~6, pp. 1378--1392, 2021.

\bibitem{cheng2018spectrally}
Z.~Cheng, B.~Liao, Z.~He, Y.~Li, and J.~Li, ``Spectrally compatible waveform
  design for mimo radar in the presence of multiple targets,'' \emph{IEEE
  Transactions on Signal Processing}, vol.~66, no.~13, pp. 3543--3555, 2018.

\bibitem{cheng2021hb}
Z.~Cheng, Z.~He, and B.~Liao, ``Hybrid beamforming for multi-carrier
  dual-function radar-communication system,'' \emph{IEEE Transactions on
  Cognitive Communications and Networking}, vol.~7, no.~3, pp. 1002--1015,
  2021.

\bibitem{liu2020joint}
X.~Liu, T.~Huang, N.~Shlezinger, Y.~Liu, J.~Zhou, and Y.~C. Eldar, ``Joint
  transmit beamforming for multiuser mimo communications and mimo radar,''
  \emph{IEEE Transactions on Signal Processing}, vol.~68, pp. 3929--3944, 2020.

\bibitem{cheng2021hybrid}
Z.~Cheng, Z.~He, and B.~Liao, ``Hybrid beamforming design for {OFDM}
  dual-function radar-communication system,'' \emph{IEEE Journal of Selected
  Topics in Signal Processing}, vol.~15, no.~6, pp. 1455--1467, 2021.

\bibitem{song2022joint}
X.~Song, D.~Zhao, H.~Hua, T.~X. Han, X.~Yang, and J.~Xu, ``Joint transmit and
  reflective beamforming for {IRS}-assisted integrated sensing and
  communication,'' in \emph{2022 IEEE Wireless Communications and Networking
  Conference (WCNC)}, 2022, pp. 189--194.

\bibitem{liu2022joint}
R.~Liu, M.~Li, Y.~Liu, Q.~Wu, and Q.~Liu, ``Joint transmit waveform and passive
  beamforming design for {RIS}-aided {DFRC} systems,'' \emph{IEEE Journal of
  Selected Topics in Signal Processing}, vol.~16, no.~5, pp. 995--1010, 2022.

\bibitem{hua2022joint}
M.~Hua, Q.~Wu, C.~He, S.~Ma, and W.~Chen, ``Joint active and passive
  beamforming design for irs-aided radar-communication,'' \emph{IEEE
  Transactions on Wireless Communications}, pp. 1--1, 2022.

\bibitem{shen2019optimization}
K.~Shen, W.~Yu, L.~Zhao, and D.~P. Palomar, ``Optimization of mimo
  device-to-device networks via matrix fractional programming: A
  minorization–maximization approach,'' \emph{IEEE/ACM Transactions on
  Networking}, vol.~27, no.~5, pp. 2164--2177, 2019.

\bibitem{wei2022joint}
T.~Wei, L.~Wu, and M.~R.~B. Shankar, ``Joint waveform and precoding design for
  coexistence of {MIMO} radar and {MU‐MISO} communication,'' \emph{IET Signal
  Processing}, vol.~16, no.~7, p. 788–799, 2022.

\bibitem{aubry2015optimizing}
A.~Aubry, A.~De~Maio, and M.~M. Naghsh, ``Optimizing radar waveform and
  {Doppler} filter bank via generalized fractional programming,'' \emph{IEEE
  Journal of Selected Topics in Signal Processing}, vol.~9, no.~8, pp.
  1387--1399, 2015.

\bibitem{grant2009cvx}
M.~Grant, S.~Boyd, and Y.~Ye, ``{CVX}: {MATLAB} software for disciplined convex
  programming,'' 2009.

\bibitem{elbir2022terahertz}
A.~M. Elbir, K.~V. Mishra, S.~Chatzinotas, and M.~Bennis, ``Terahertz-band
  integrated sensing and communications: {C}hallenges and opportunities,''
  \emph{arXiv preprint arXiv:2208.01235}, 2022.

\end{thebibliography}

\ifCLASSOPTIONcaptionsoff
  \newpage
\fi

\end{document}